\RequirePackage{fix-cm}
\documentclass[smallextended,natbib]{svjour3}       
\smartqed  
\usepackage{graphicx}
\usepackage{mathptmx}      
\usepackage{amsmath}
\usepackage{amsfonts}
\usepackage{amssymb}

\newcommand{\deriv}[3][]{\ensuremath{\frac{\mathrm{d}^{#1}#2}{\mathrm{d}#3^{#1}}}}

\renewcommand{\arctan}{\mathrm{tan}^{-1}}

\journalname{Celestial Mechanics and Dynamical Astronomy}

\usepackage[utf8]{inputenc}
\usepackage{color}

\begin{document}

\title{Multiple Scales Asymptotic Solution For The Constant Radial Thrust Problem}

\author{Juan Luis Gonzalo \and Claudio Bombardelli }

\institute{J. L. Gonzalo \at Postdoctoral research fellow. Department of Aerospace Science and Technology, Politecnico di Milano, Via la Masa 34, Milan 20156, Italy\\
E-mail: juanluis.gonzalo@polimi.it\\
ORCID: 0000-0002-2181-6303
\and
C. Bombardelli \at Associate Professor. Space Dynamics Group, School of Aerospace Engineering (ETSIAE), Technical University of Madrid-UPM, Plaza Cardenal Cisneros 3, Madrid 28040, Spain
}

\date{Received: date / Accepted: date}

\maketitle

\begin{abstract}

An approximate analytical solution for the two body problem perturbed by a radial, low acceleration is obtained, using a regularized formulation of the orbital motion and the method of multiple scales. Formulating the dynamics with the Dromo special perturbation method allows us to separate the two characteristic periods of the problem in a clear and physically significative way, namely the orbital period and a period depending on the magnitude of the perturbing acceleration. This second period becomes very large compared to the orbital one for low thrust cases, allowing us to develop an accurate approximate analytical solution through the method of multiple scales. Compared to a regular expansion, the multiple scales solution retains the qualitative contributions of both characteristic periods and has a longer validity range in time. Looking at previous solutions for this problem, our approach has the advantage of not requiring the evaluation of special functions or an initially circular orbit. Furthermore, the simple expression reached for the long period provides additional insight on the problem. Finally, the behavior of the asymptotic solution is assessed through several test cases, finding a good agreement with high-precision numerical solutions. The results presented not only advance in the study of the two body problem with constant radial thrust, but confirm the utility of the method of multiple scales for tackling problems in orbital mechanics.

\end{abstract}

\section{Introduction}\label{sec:introduction}

The method of multiple scales [see \cite{bender2013advanced,murdock1999perturbations,Nayfeh,Hinch}] is a powerful perturbation technique that can be employed to study the behavior of complex dynamical systems. It is particularly effective when secular terms are introduced in the dynamics causing classical perturbation methods to diverge. In such circumstance, the approximate analytical solution obtained with a straightforward (regular) expansion is only valid within a small interval which decreases as the small perturbing parameter ($\varepsilon$) grows.

Multiple scales methods have been widely employed in almost all branches of physics, ranging from molecular dynamics to computational fluid mechanics. However, their use in orbital mechanics is relatively scarce and limited to a few references; see for instance \cite{kevorkian1987perturbation}, page 396 of 391-461, or \cite{karlgaard2003second}. This contrasts with the common application in orbital mechanics of other perturbation techniques such as averaging or variation of the parameters. Even other aymptotic perturbation methods are common in the literature. In \cite{zuiani2012direct}, a first order approximate analytical solution of Gauss planetary equations is exploited for the direct optimization of low-thrust trajectories. The same description of the dynamics is applied by \cite{avanzini2015solution} in order to deal with the low-thrust Lambert problem. Asymptotic series expansion solutions for the constant tangential and circumferential acceleration problems are given in \cite{bombardelli2011asymptotic} and \cite{niccolai2018orbital}, respectively.

One interesting orbital mechanics problem that can be tackled with the multiple scales perturbation technique is that of a body orbiting a primary and perturbed by a constant radial acceleration. The problem was first investigated by \cite{tsien1953take}. In the classic astrodynamics book of \cite{battin1999introduction} an exact solution for the case of initially circular orbit is reached in terms of elliptic integrals. Several authors have since dealt with the problem and its nuances: \cite{quarta2012new,akella2002anatomy,prussing1998constant,san2012bounded,calvo2019Approximate}. 
\cite{quarta2012new} noted that, although several exact results were known in terms of elliptic functions, the lack of physical insight limited their practical use for mission design. To tackle this, they introduced a regularization of the equations of motion and obtained an implicit solution for the trajectory in terms of an infinite Fourier series. Subsequently, they derived an explicit, approximate solution based on the implicit, exact one. \cite{calvo2019Approximate} build on this approach by proposing a new procedure to compute the coefficients of the approximate Fourier series. A Hamiltonian-based solution is given by \cite{san2012bounded} for the bounded case, characterizing the different regimes as the result of a bifurcation phenomenon. Differently from \cite{quarta2012new}, they face the lack of physical insight by providing a qualitative description of the flow in the energy-momentum plane. More recently, \cite{urrutxua2015dromo} proposed a new exact solution based on a regularized orbital formulation using elliptical integral functions. However, their solution is still restricted to initially circular orbit. An exact solution for the radial problem in the general case was recently obtained by \cite{izzo2014explicit}, in terms of a fictitious time introduced with a Sundman transformation and the Weierstrass elliptic functions. However, the undoubtedly elegant solution involves elliptic functions, with the physical interpretation and analytic manipulation difficulties noted by other authors [\cite{quarta2012new,san2012bounded}].

In the present work the use of the multiple scales perturbation method is exploited in order to obtain an approximate solution with a simpler analytical representation but retaining high accuracy across a reasonably wide range of initial conditions and acceleration magnitudes. As it will be shown in the results the approximate solution obtained can be expressed in terms of simpler functions providing more insight about the underlying physics of the problem. In particular, relatively compact expressions are provided to quickly evaluate the main frequencies of the dynamics which are not straightforward to compute even using numerical integration. Last but not least, the article provides a relatively simple example of the application of the multiple scale technique to an example of perturbed two-body problem.

One fundamental aspect of the analytical procedure employed in the article is the use of a convenient orbital motion formulation to ease the analytical treatment as much as possible. The so called ``Dromo'' formulation initially introduced by \cite{Pelaez} and complemented by other authors [see \cite{Bau2013,bau2014time}] is chosen here due to its suitability when dealing with a constant radial thrust problem. This is because it is intrinsically based on a local vertical- local horizontal orbiting frame where a purely radial acceleration component appears as a simple perturbing term in the equations of motions. These advantages have been exploited in the solution to the radial thrust problem proposed by \cite{urrutxua2015dromo}. Dromo has also proven to be an excellent propagation tool, and its suitability for the formulation of low thrust optimal control problems has been recently studied by \cite{Gonzalo2012PFC,gonzalo2015low}. Also, an asymptotic solution for low tangential acceleration was obtained by \cite{bombardelli2011asymptotic} based on the same formulation and using a regular expansion in the non-dimensional thrust.

The article is structured as follows. First, the equations of motion for the constant radial thrust problem are written in Dromo variables and an asymptotic solution for the problem is obtained using a regular expansion in the small perturbing acceleration. It is seen that this solution breaks for large values of the independent variable, suggesting the use of more complex perturbation techniques. The following section deals with the solution of the problem using the method of multiple scales. Then, the multiple scales solution is compared with high-precision numerical propagations for several cases, finding a good agreement between them. The asymptotic solution obtained through the regular expansion is also included in these comparisons, highlighting the great gains in accuracy and validity range of the asymptotic solution achieved through the method of multiple scales. Finally, a conceptual comparison with other methods is presented, and conclusions are drawn.

\section{Equations of Motion}

Let us consider a particle orbiting around a primary of gravitational constant $\mu$ and perturbed by a radial acceleration of constant magnitude $A$, leading to a planar motion with constant angular momentum. Let $R_{0}$ denote the initial distance between the particle and the center of the primary and $\nu_{0}$ the initial value of the true anomaly. All equations and variables considered hereafter are expressed in non-dimensional form taking $R_{0}$ as characteristic length and $1/n_{0}$ as characteristic time, where $n_{0}=(\mu/R_{0}^3)^{1/2}$ is the angular frequency of the circular orbit of radius $R_{0}$ around the primary.

To describe the motion of the body, the Dromo orbital formulation developed by \cite{Pelaez} is used. In this formulation a fictitious time $\theta$ is introduced through a change of independent variable given by a Sundman's transformation. Then, the variation of parameters technique is applied to obtain seven generalized orbital parameters $\mathbf{q}$ which, along with the non-dimensional time $t$, describe the state of the particle. These orbital parameters are constant in the unperturbed problem, but evolve in presence of perturbing forces; this property is very convenient for the mathematical developments in this study. Moreover, three of these parameters describe the geometry of the orbit in its plane, while the other four are related to the orientation of said plane. Therefore, the latter are constant for the planar case, and the motion of the particle  can be described by a 4-dimensional state vector: 
\begin{equation}
(t,q_{1},q_{2},q_{3}) \, ,
\end{equation}
whose evolution for the radial thrust problem with a perturbing acceleration of magnitude $A$ is given by the following system of four differential equations: 
\begin{equation}\label{eq:Dromo_ODEs_time}
\frac{\mathrm{d}t}{\mathrm{d}\theta}=\frac{1}{q_{3}s^{2}} \, ,
\end{equation}
\begin{equation}\label{eq:radial_Dromoeq}
\frac{\mathrm{d}}{\mathrm{d}\theta}\left[\begin{array}{c}
q_{1}\\
q_{2}\\
q_{3}
\end{array}\right]=\frac{\varepsilon}{q_{3}s^{2}}\left[\begin{array}{c}
\sin\theta\\
-\cos\theta\\
0
\end{array}\right] \, ,
\end{equation}
with initial conditions:
\[
t(\theta_{0})=t_{i} \, , \qquad q_{1}(\theta_{0})=q_{1i} \, , \qquad q_{2}(\theta_{0})=q_{2i} \, , \qquad q_{3}(\theta_{0})=q_{3i} \, ,
\]
where $\varepsilon$ is the non-dimensional acceleration parameter:
\begin{equation}
\varepsilon=\frac{A}{R_{0}n_{0}^{2}}=\frac{A}{\mu/R_{0}^{2}} \, ,
\end{equation}
and $s$ corresponds to the non-dimensional velocity in the transversal direction:
\begin{equation}\label{eq:Dromo_s}
s=q_{3}+q_{1}\cos\theta+q_{2}\sin\theta \, .
\end{equation}
The non-dimensional orbital distance can be related to previous quantities by the relation:
\begin{equation}\label{eq:Dromo_r}
r=\frac{1}{q_{3}s}\, .
\end{equation}

It is also  possible to establish several relations between the generalized orbital parameters and the classical orbital elements, as developed by \cite{Urrutxua2013} and \cite{Bau2013}: 
\begin{equation}\label{eq:Dromo_fromclassic}
q_{1}=\frac{e}{h}\cos\gamma \, ,\qquad q_{2}=\frac{e}{h}\sin\gamma \, ,\qquad q_{3}=\frac{1}{h} \, ,
\end{equation}
\begin{equation}\label{eq:Dromo_fromDromo1}
e=\frac{\sqrt{q_{1}^{2}+q_{2}^{2}}}{q_{3}} \, ,\quad\gamma=\tan^{-1}\left(\frac{q_{2}}{q_{1}}\right) \, ,\quad h=\frac{1}{q_{3}} \, ,
\end{equation}
\begin{equation}\label{eq:Dromo_fromDromo2}
a=\frac{1}{q_{3}^{2}-q_{1}^{2}-q_{2}^{2}} \, ,\quad E=\frac{q_{1}^{2}+q_{2}^{2}-q_{3}^{2}}{2} \, ,\quad E_{\text{T}}=E-\varepsilon r \, .
\end{equation}
In the above expressions, $e$ is the eccentricity, $h$ is the non-dimensional angular momentum, $a$ is the non-dimensional semimayor axis, $E$ is the non-dimensional Keplerian energy, $E_{T}$ is the non-dimensional total energy, and the angle $\gamma$ is the difference between the variations of the fictitious time and the true anomaly [see \cite{Bau2013}]:
\begin{equation}\label{eq:Dromo_nu_theta_gamma}
\Delta\theta=\Delta\nu+\gamma \, .
\end{equation}
In the planar case, $\gamma$ coincides with the angle between the eccentricity vectors of the initial and of the osculating orbit, and $\theta$ becomes the angular position of the particle measured from the eccentricity vector of the initial orbit.

To close the mathematical formulation of the dynamics the initial conditions are obtained from Eq.~\eqref{eq:Dromo_fromclassic}, taking into account that $\theta_{0}=\nu_{0}$, $\gamma_{0}=0$: 
\begin{equation}\label{eq:radial_Dromoic}
q_{1i}=\frac{e_{0}}{h_{0}} \, , \quad q_{2i}=0 \, , \quad q_{3i}=\frac{1}{h_{0}} \, ,
\end{equation}
with $h_{0}=\sqrt{1+e_{0}\cos\nu_{0}}$. For simplicity and clarity, in the following developments the initial value of the independent variable is assumed to be zero, $\theta_{0}=0$. The results can be generalized for an arbitrary value of $\theta_{0}$ by introducing the corresponding integration constants.

While the generalized orbital element $q_{3}$ is constant as a consequence of the conservation of angular momentum, the other two vary according to Eq.~\eqref{eq:radial_Dromoeq} whose solution will be approached in the remainder of the article.

\section{Regular expansion}

\subsection{Generalized orbital elements $q_1$, $q_2$}

An asymptotic solution for the low-thrust two body problem defined by {Eqs.~(\ref{eq:radial_Dromoeq},\ref{eq:radial_Dromoic})} is now sought for, in the form of a regular expansion\footnote{For clarity, the symbol $\, \hat{} \,$ will be used to denote all the variables related to this asymptotic solution.} in the non-dimensional thrust parameter $\varepsilon$. To this end, the state is expanded in power series of $\varepsilon\ll1$ as follows:
\[
\begin{array}{c}
\hat{q}_{1}(\theta;\varepsilon)=\hat{q}_{10}(\theta)+\varepsilon\hat{q}_{11}(\theta)+\varepsilon^{2}\hat{q}_{12}(\theta)+\mathcal{O}(\varepsilon^{3}) \, ,\\
\hat{q}_{2}(\theta;\varepsilon)=\hat{q}_{20}(\theta)+\varepsilon\hat{q}_{21}(\theta)+\varepsilon^{2}\hat{q}_{22}(\theta)+\mathcal{O}(\varepsilon^{3}) \, .
\end{array}
\]
Expanding also the initial conditions, with $\theta_{0}=0$, and identifying terms of equal power of $\varepsilon$:
\begin{equation}
\begin{array}{c}
\hat{q}_{10}(0)=q_{1i},\qquad\hat{q}_{11}(0)=\hat{q}_{12}(0)=0 \, ,\\
\hat{q}_{20}(0)=\hat{q}_{21}(0)=\hat{q}_{22}(0)=0 \, .
\end{array}\label{eq:radial_ic}
\end{equation}
For compactness, the power series expansion of $\hat{s}$ is
also defined as:
\[
\hat{s}(\theta)=\hat{s}_{0}(\theta)+\varepsilon\hat{s}_{1}(\theta)+\mathcal{O}\left(\varepsilon^{2}\right) \, ,
\]
with
\[
\begin{array}{c}
\hat{s}_{0}(\theta)=q_{3i}+\hat{q}_{10}\cos\theta+\hat{q}_{20}\sin\theta \, ,\\
\hat{s}_{1}(\theta)=\hat{q}_{11}\cos\theta+\hat{q}_{21}\sin\theta \, .
\end{array}
\]

Introducing the expansion of the state into the first two components of Eq.~\eqref{eq:radial_Dromoeq}, expanding in Taylor series of $\varepsilon$ and retaining the leading order terms yields:
\[
\frac{\mathrm{d}}{\mathrm{d}\theta}\left[\begin{array}{c}
\hat{q}_{10}\\
\hat{q}_{20}
\end{array}\right]=\left[\begin{array}{c}
0\\
0
\end{array}\right]\quad\Rightarrow\quad\left[\begin{array}{c}
\hat{q}_{10}\\
\hat{q}_{20}
\end{array}\right]=\left[\begin{array}{c}
q_{1i}\\
0
\end{array}\right] \, .
\]
That is, the zeroth order terms of the asymptotic solution are constant and equal to their initial values; this result was expected, since the limit $\varepsilon=0$ corresponds to the unperturbed orbit, for which Dromo orbital parameters remain constant. Additionally, the expression for $\hat{s}_{0}$ now takes the simpler form: 
\begin{equation}
\hat{s}_{0}=q_{3i}+q_{1i}\cos\theta \, .
\end{equation}

To retain the effect of the thrust, it is necessary to consider the first order terms of the asymptotic solution. The differential equations describing their evolution with $\theta$ are obtained canceling terms of $\mathcal{O}(\varepsilon)$ in the expansion of Eq.~\eqref{eq:radial_Dromoeq}: 
\begin{equation}\label{eq:radial_Dromoeq_Oe}
\frac{\mathrm{d}}{\mathrm{d}\theta}\left[\begin{array}{c}
\hat{q}_{11}\\
\hat{q}_{21}
\end{array}\right]=\frac{1}{q_{3i}\left(\hat{s}_{0}\right)^{2}}\left[\begin{array}{c}
\sin\theta\\
-\cos\theta
\end{array}\right] \, .
\end{equation}
Since the previous equations are uncoupled, $\hat{q}_{11}(\theta)$ and $\hat{q}_{21}(\theta)$ can be obtained independently as quadratures. Introducing the known results for $\hat{q}_{10}$ and $\hat{q}_{20}$, and taking into account the initial conditions given by Eq.~\eqref{eq:radial_ic}, the solution for $\hat{q}_{11}(\theta)$ is straightforward:
\begin{equation}
\hat{q}_{11}(\theta)=\frac{1-\cos\theta}{q_{3i}(q_{3i}+q_{1i})\hat{s}_{0}} \, .
\end{equation}

The solution for $\hat{q}_{21}(\theta)$ is more complex. Assuming $q_{3i}>q_{1i}$ and integrating\footnote{This assumption is valid as long as $e_{0}<1$, since $q_{1i}=e_{0}q_{3i}$.} yields: 
\begin{equation}
\hat{q}_{21}(\theta)=\frac{-\sin\theta}{\left(q_{3i}^{2}-q_{1i}^{2}\right)\hat{s}_{0}}+\frac{2q_{1i}}{q_{3i}\left(q_{3i}^{2}-q_{1i}^{2}\right)^{3/2}}\left(\frac{\theta}{2}+\tan^{-1}\hat{\mathcal{K}}\right) \, ,
\end{equation}
with:
\[
\hat{\mathcal{K}}=-\frac{\sin\theta\left(-q_{3i}+q_{1i}+\sqrt{q_{3i}^{2}-q_{1i}^{2}}\right)}{(1+\cos\theta)\sqrt{q_{3i}^{2}-q_{1i}^{2}}+(1-\cos\theta)(q_{3i}-q_{1i})} \, .
\]
The second term of the expression for $\hat{q}_{21}$ introduces a secular behavior in $\theta$; as a result, $\varepsilon\hat{q}_{21}$ becomes of order one for $\theta\sim1/\varepsilon$ and the regular asymptotic expansion breaks down. This posses a clear limitation to the applicability of this solution for long term propagation, and suggests the convenience of resorting to more complex formulations.

\subsection{Physical Time}

The previous results are given in the fictitious time introduced by the Sundman transformation, which in this case coincides with the angular position measured from the initial eccentricity vector. Therefore, it is interesting to establish a relation between this fictitious time $\theta$ and the non-dimensional physical time $\hat{t}$. However, instead of working directly with the physical time and Eq.~\eqref{eq:Dromo_ODEs_time}, the time element for the Dromo formulation proposed by \cite{bau2014time} is preferred. Particularizing their equations for the constant radial thrust case, the algebraic relation between the physical time and the time element $\zeta_{0}$ takes the form:
\begin{equation}
\zeta_{0}=t-\frac{u}{2Eq_{3}s}-\frac{1}{E\sqrt{-2E}}\arctan\left(\frac{u}{s+\sqrt{-2E}}\right)\, ,\label{eq:Dromo_TE_t_relation}
\end{equation}
and the differential equation describing the evolution of said time element with $\theta$ is:
\begin{equation}\label{eq:Dromo_TE_t_ODE}
\deriv{\zeta_{0}}{\theta}=a^{3/2}\left[1+\frac{\varepsilon u}{q_{3}s^{2}}\left(6a\arctan\left(\frac{u}{f+w}\right)+k_{1}\right)+\frac{\varepsilon}{q_{3}s}k_{2}\right] \, ,
\end{equation}
where:
\[
k_{1}=\frac{\sqrt{a}u}{s^{2}}\left(\frac{q_{3}+s}{f}+\frac{2w}{q_{3}}+1\right)\,,\quad k_{2}=\frac{1}{s^{2}}\left(\frac{f}{q_{3}}+\frac{w}{f}+\frac{u^{2}}{fs}\right) \, ,
\]
\[
u=q_{1}\sin\theta-q_{2}\cos\theta\,,\quad w=q_{1}\cos\theta+q_{2}\sin\theta \, ,
\]
\[
f=q_{3}+\sqrt{-2E} \, .
\]
Denoting with $\hat{\zeta}_0$ the time element for the regular asymptotic solution, and expanding it in power series of $\varepsilon$:
\[
\hat{\zeta}_0 \left( \theta \right) = \hat{\zeta}_{00} \left( \theta \right) + \varepsilon \hat{\zeta}_{01} \left( \theta \right) + \varepsilon^2 \hat{\zeta}_{02} \left( \theta \right) + \mathcal{O} \left( \varepsilon^3 \right) \, .
\]
The zeroth-order term $\hat{\zeta}_{00}$ can be obtained by introducing this expression and the known solutions for $\hat{q}_1$ and $\hat{q}_2$ into Eq.~\eqref{eq:Dromo_TE_t_ODE}, expanding in power series of $\varepsilon$ up to the leading order terms and integrating:
\begin{equation}\label{eq:reg_time_el}
\hat{\zeta}_{00} = \frac{1}{\left( q_{3i} - q_{1i} \right)^{3/2}} \theta \, ,
\end{equation}
which is strictly secular in $\theta$, with no oscillatory terms.

The asymptotic solution for the physical time $\hat{t}$ is recovered by substituting the expressions for $\hat{\zeta}_0$, $\hat{q}_1$, and $\hat{q}_2$ back into Eq.~\eqref{eq:Dromo_TE_t_relation}. It includes both secular and oscillatory terms in $\theta$, which is consistent as time must be monotonically increasing with $\theta$. However, although one would expect for the secular growth to affect only the time element this is not the case, because of the undesired secular term in $\hat{q}_2$. This not only contributes to the breakdown of the solution for $\theta \sim 1/\varepsilon$, but Eq.~\eqref{eq:Dromo_TE_t_relation} cannot even be evaluated to real values when the spurious secular term in $\hat{q}_2$ leads to positive $E$. Nevertheless, this latter issue does not introduce additional restrictions to the validity range of the solution as it is just a side effect of the normal breakdown of the regular expansion. For completeness, the expression for $\hat{t}$ obtained directly from Eq.~\eqref{eq:Dromo_ODEs_time}, which always evaluates to real values, is reported in Appendix~\ref{ap:time_regular}.

\subsection{Orbital Elements}

The asymptotic solution is given in terms of Dromo generalized parameters, but it is not straightforward to give physical interpretations for all of them. Therefore, it is convenient to express it in terms of more familiar orbital elements, by substituting $\hat{q}_{1}(\theta;\varepsilon)$ and $\hat{q}_{2}(\theta;\varepsilon)$ into {Eqs.~(\ref{eq:Dromo_fromDromo1}-\ref{eq:Dromo_fromDromo2})} and expanding in power series of $\varepsilon$ up to first order. Proceeding in this manner, the following expression for the eccentricity $\hat{e}$ is obtained: 
\begin{equation}\label{eq:reg_e_exp}
\hat{e}(\theta;\varepsilon)=e_{0}+\varepsilon\,\frac{1-\cos\theta}{q_{3i}^{4}(1+e_{0})(1+e_{0}\cos\theta)}+\mathcal{O}(\varepsilon^{2}) \, .
\end{equation}
As a first approximation, the eccentricity oscillates between $e_{0}$ and $e_{0}+2\varepsilon q_{3i}^{-4}(1-e_{0}^{2})^{-1}$ with a period of $2\pi$ in the fictitious time $\theta$. Note that for $e_{0}/\varepsilon \lesssim 1$ the preceding expansion ceases to be valid, and should be replaced by the full Eq.~\eqref{eq:Dromo_fromDromo1} or by the expression for initially circular or quasi-circular orbit derived later in this article.

Likewise, the expansion for the orbit radius (valid regardless of the value of $e_{0}$) yields:
\begin{equation}
\hat{r}(\theta;\varepsilon)=\frac{1}{q_{3i}\hat{s}_{0}}+\varepsilon\,\frac{-q_{1i}\sin\theta\left(\theta+2\arctan\hat{\mathcal{K}}\right)+(1-\cos\theta)\sqrt{q_{3i}^{2}-q_{1i}^{2}}}{q_{3i}^{2}(q_{3i}^{2}-q_{1i}^{2})^{3/2}\left(\hat{s}_{0}\right)^{2}}+\mathcal{O}(\varepsilon^{2}).
\end{equation}
From a physical point of view, the presence of the secular term in $\theta$ would imply that the orbital radius is unbounded for any value of $e_{0}$ and $\varepsilon$, which is in contradiction with the results given by \cite{battin1999introduction}. Moreover, the mathematical validity of the expansion breaks for $\theta \sim 1/\varepsilon$. On the other hand, the leading order term of $\hat{r}$ turns out to be $2\pi$ periodic in $\theta$, retaining the same period as the unperturbed motion.

The Keplerian energy expansion yields:
\begin{equation}
\hat{E}\left(\theta;\varepsilon\right)=\frac{q_{1i}^{2}-q_{3i}^{2}}{2}+\varepsilon\frac{q_{1i}\left(1-\cos\theta\right)}{q_{3i}(q_{3i}+q_{1i})\left(q_{3i}+q_{1i}\cos\theta\right)}+\mathcal{O}\left(\varepsilon^{2}\right) \, ,
\end{equation}
corresponding to small oscillations about the initial value. On the other hand, the total energy $\hat{E}_{\text{T}}$ expansion yields the constant value (satisfying energy conservation): 
\begin{equation}
\hat{E}_{\text{T}}(\theta;\varepsilon)=\frac{q_{1i}^{2}-q_{3i}^{2}}{2}-\varepsilon\frac{1}{q_{3i}\left(q_{3i}+q_{1i}\right)}+\mathcal{O}\left(\varepsilon^{2}\right) \, .
\end{equation}

Finally, it is interesting to consider the evolution of the angle $\hat{\gamma}$ formed by the initial and osculating eccentricity vectors, which is also the angular drift between the true anomaly $\nu$ and the fictitious time $\theta$. Following Eq.~\eqref{eq:Dromo_fromDromo1}, the regular expansion of $\hat{\gamma}$ up to $\mathcal{O}(\varepsilon)$ can be given as: 
\begin{equation}
\hat{\gamma}(\theta;\varepsilon)=-\varepsilon\left(\frac{\sin\theta}{q_{1i}(q_{3i}^{2}-q_{1i}^{2})\left(q_{3i}+q_{1i}\cos\theta\right)}-\frac{\theta+2\arctan\hat{\mathcal{K}}}{q_{3i}(q_{3i}^{2}-q_{1i}^{2})^{3/2}}\right)+\mathcal{O}\left(\varepsilon^{2}\right).
\end{equation}
Same as with $\hat{r}$, this expression includes a secular behavior in $\theta$. However, the lack of a zeroth order term introduces the reasonable doubt of whether this is just a mathematical artifact or it actually represents a physical characteristic of the solution. This question shall be addressed in the following section, where a more accurate asymptotic solution is reached. Aside from this, the previous expansion is not valid for the case of initially circular or quasi-circular orbits; in those cases, the general expression in Eq.~\eqref{eq:Dromo_fromDromo1} must be used.

\subsection{Initially circular orbit}

For the particular case of an orbit with $e_{0}=0$, the regular asymptotic solution presented in this section takes a simpler form: 
\[
\hat{q}_{1}(\theta;\varepsilon)=\varepsilon\frac{1-\cos\theta}{q_{3i}^{3}}+\mathcal{O}(\varepsilon^{2}) \, , \quad \hat{q}_{2}(\theta;\varepsilon)=\varepsilon\frac{-\sin\theta}{q_{3i}^{3}}+\mathcal{O}(\varepsilon^{2}) \, ,
\]
and from it: 
\[
\hat{e}(\theta;\varepsilon)=\varepsilon\frac{\sqrt{2}}{q_{3i}^{4}}\sqrt{1-\cos\theta}+\mathcal{O}(\varepsilon^{2}) \, , \quad \hat{r}(\theta;\varepsilon)=\frac{1}{q_{3i}^{2}}+\varepsilon\frac{1-\cos\theta}{q_{3i}^{6}}+\mathcal{O}(\varepsilon^{2}) \, ,
\]
\[
\hat{E}(\theta;\varepsilon)=-\frac{q_{3i}^{2}}{2}+\mathcal{O}\left(\varepsilon^{2}\right) \, , \quad \hat{E}_{\text{T}}(\theta;\varepsilon)=-\frac{q_{3i}^{2}}{2}-\varepsilon\frac{1}{q_{3i}^{2}}+\mathcal{O}\left(\varepsilon^{2}\right) \, ,
\]
\[
\hat{\gamma}(\theta;\varepsilon)=\frac{\theta-\pi}{2}+\mathcal{O}(\varepsilon^{2}) \, .
\]
The secular term in $\theta$ has vanished from the solutions for $\hat{q}_{2}$ and $\hat{r}$, but this does not imply a good behavior for large values of the independent variable $\theta$. Certainly, the numerical results displayed in later sections show that the approximation is still bad for $\theta \sim 1/\varepsilon$. It is also observed that the energy oscillates comparatively less than the other elements, no longer containing a term of $\mathcal{O}\left(\varepsilon\right)$.

\section{Multiple Scales}\label{sec:MultipleScales}

\subsection{Generalized orbital elements $q_{1}$, $q_{2}$}

The breakdown of the regular expansion for $\theta\sim1/\varepsilon$ suggests the existence of a slow `time' scale in the independent variable $\theta$. This hypothesis is further supported by the perturbation model for the tangential case presented by \cite{bombardelli2011asymptotic}, and the solutions in terms of elliptic equations given for the radial  thrust problem by \cite{battin1999introduction}. All those cases are characterized by a fast, periodic evolution associated to the orbital period, and a slow, secular behavior whose characteristic period depends on the magnitude of the thrust. Furthermore, previous works by the authors [see \cite{Gonzalo2014}] confirm that the mathematical structure of the multiple scales problem supports exactly two independent `time' scales, with additional slower scales being a correction of the slow one. Therefore, a high order multiple scales solution is proposed by introducing a fast `time' scale $\tau$ and a slow `time' scale $T$:
\[
\tau=\theta \, ,\qquad T=\Omega(\varepsilon)\,\theta \, ,
\]
where the unknown function $\Omega(\varepsilon)$ represents a coordinate strain in the slow `time' scale $T$. Note that it must fulfill $\Omega(0)=0$. Using the chain rule, the derivative operator can be rewritten in terms of the new independent variables as: 
\[
\deriv{}{\theta}=\frac{\partial}{\partial\tau}\deriv{\tau}{\theta}+\frac{\partial}{\partial T}\deriv{T}{\theta}=\frac{\partial}{\partial\tau}+\Omega(\varepsilon)\frac{\partial}{\partial T} \, ,
\]
and introducing it into Eq.~\eqref{eq:radial_Dromoeq} yields:
\begin{equation}\label{eq:Dromo_radial_ms_q1q2}
\frac{\partial}{\partial\tau}\left[\begin{array}{c}
q_{1}(\tau,T)\\
q_{2}(\tau,T)
\end{array}\right]+\Omega(\varepsilon)\frac{\partial}{\partial T}\left[\begin{array}{c}
q_{1}(\tau,T)\\
q_{2}(\tau,T)
\end{array}\right]=\frac{\varepsilon}{q_{3}s^{2}}\left[\begin{array}{c}
\sin\tau\\
-\cos\tau
\end{array}\right] \, .
\end{equation}
The unknown function giving the coordinate strain in the slow scale will be approximated through its power series expansion in $\varepsilon$:
\begin{equation}\label{eq:radial_ms_Omegaexp}
\Omega(\varepsilon)=\Omega_{1}\varepsilon\left(1+\Omega_{2}\varepsilon\right)+\mathcal{O}(\varepsilon^{3}) \, .
\end{equation}
Note that coefficient $\Omega_{1}$ will not be determined by the secularity conditions in the multiple scales problem; it has been included as an additional degree of freedom and its value will be chosen as to simplify the expressions obtained for the zeroth order solution.

The series expansions for $q_{1}(\tau,T)$ and $q_{2}(\tau,T)$ in $\varepsilon\ll1$ up to second order terms are now of the form:
\begin{equation}\label{eq:radial_ms_qexp}
\begin{array}{c}
q_{1}(\tau,T;\varepsilon)=q_{10}(\tau,T)+\varepsilon q_{11}(\tau,T)+\varepsilon^{2}q_{12}(\tau,T)+\mathcal{O}(\varepsilon^{3}) \, ,\\
q_{2}(\tau,T;\varepsilon)=q_{20}(\tau,T)+\varepsilon q_{21}(\tau,T)+\varepsilon^{2}q_{22}(\tau,T)+\mathcal{O}(\varepsilon^{3}) \, ,
\end{array}
\end{equation}
with initial conditions:
\begin{equation}\label{eq:radial_ms_ic}
\begin{array}{c}
q_{10}(0,0)=q_{1i},\qquad q_{11}(0,0)=q_{12}(0,0)=0 \, ,\\
q_{20}(0,0)=q_{21}(0,0)=q_{22}(0,0)=0 \, .
\end{array}
\end{equation}
We also define:
\begin{equation}
s(\tau,T;\varepsilon)=s_{0}(\tau,T)+\varepsilon s_{1}(\tau,T)+\mathcal{O}\left(\varepsilon^{2}\right) \, ,
\end{equation}
with:
\[
s_{0}(\tau,T)=q_{3i}+q_{10}\cos\tau+q_{20}\sin\tau,
\]
\[
s_{1}(\tau,T)=q_{11}\cos\tau+q_{21}\sin\tau.
\]

Substituting Eq.~\eqref{eq:radial_ms_qexp} into Eq.~\eqref{eq:Dromo_radial_ms_q1q2}, expanding in power series of the small parameter $\varepsilon$ and retaining terms of order $\mathcal{O}(\varepsilon^{0})$ yields: 
\begin{equation*}
\frac{\partial}{\partial\tau}\left[\begin{array}{c}
q_{10}\\
q_{20}
\end{array}\right]=\left[\begin{array}{c}
0\\
0
\end{array}\right]\Rightarrow\left[\begin{array}{c}
q_{10}(\tau,T)\\
q_{20}(\tau,T)
\end{array}\right]=\left[\begin{array}{c}
q_{10}(T)\\
q_{20}(T)
\end{array}\right] \, ,
\end{equation*}
with:
\begin{equation*}
q_{10}(0)=q_{1i} \, , \quad  q_{20}(0)=0 \, .
\end{equation*}
Hence, $q_{10}$ and $q_{20}$ are no longer constants, unlike their regular expansion counterparts $\hat{q}_{10}$ and $\hat{q}_{20}$, but rather functions of the slow scale $T$. Moreover, $q_{20}(T)\neq0$ in general, and the corresponding simplifications introduced in the derivation of the regular solution cannot be made.

The zeroth order equations do not provide enough information to fully determine $q_{10}$ and $q_{20}$. This is the expected behavior when applying the method of multiple scales. As shown in many classic perturbation texts [see \cite{Nayfeh,Hinch,murdock1999perturbations,bender2013advanced}], the equations to close the zeroth order solution will be obtained from the cancellation of the secular terms in the subsequent order solution (also known as \textit{secularity condition}). Collecting terms of order $\mathcal{O}(\varepsilon)$ in the expansion of Eq.~\eqref{eq:Dromo_radial_ms_q1q2} yields:
\begin{equation*}
\frac{\partial}{\partial\tau}\left[\begin{array}{c}
q_{11}\\
q_{21}
\end{array}\right]+\Omega_{1}\frac{\partial}{\partial T}\left[\begin{array}{c}
q_{10}\\
q_{20}
\end{array}\right]=\frac{1}{q_{3i}s_{0}^{2}}\left[\begin{array}{c}
\sin\tau\\
\cos\tau
\end{array}\right] \, .
\end{equation*}
Rearranging terms and integrating with respect to $\tau$, the first
order solutions $q_{11}$ and $q_{21}$ are reached:
\begin{equation}
\left[\begin{array}{c}
q_{11}\\
q_{21}
\end{array}\right]=\left[\begin{array}{c}
\mathcal{P}_{1}\\
\mathcal{P}_{2}
\end{array}\right]+\left(\frac{\tau}{2}+\tan^{-1}\mathcal{K}\right)\left[\begin{array}{c}
\mathcal{S}_{1}\\
\mathcal{S}_{2}
\end{array}\right]-\tau\Omega_{1}\frac{\partial}{\partial T}\left[\begin{array}{c}
q_{10}\\
q_{20}
\end{array}\right]+\left[\begin{array}{c}
g_{1}\\
g_{2}
\end{array}\right].\label{eq:MTS_sol}
\end{equation}
In the above expressions, the functions $\mathcal{P}_{1}\left(\tau,T\right)$, $\mathcal{P}_{2}\left(\tau,T\right)$, $\mathcal{K}\left(\tau,T\right)$, $\mathcal{S}_{1}\left(T\right)$ and $\mathcal{S}_{2}\left(T\right)$ are reported in Appendix \ref{ap:ms_components}. Suffice to say that they are all periodic functions of the fast scale $\tau$ while their dependence on the slow `time' scale $T$ only appears through the zeroth order terms $q_{10}$ and $q_{20}$. In addition, two unknown functions of the slow scale, $g_{1}\left(T\right)$ and $g_{2}\left(T\right)$, appear as constants of integration.

Imposing the cancellation of the $\tau$ secular terms in Eq.~\eqref{eq:MTS_sol} yields the following ODE system for $q_{10}$ and $q_{20}$: 
\begin{equation}\label{eq:Dromo_eq_q10T}
\Omega_{1}\deriv{q_{10}}{T}=-\frac{q_{20}}{q_{3i}\left(q_{3i}^{2}-q_{10}^{2}-q_{20}^{2}\right)^{3/2}} \, ,
\end{equation}
\begin{equation}\label{eq:Dromo_eq_q20T}
\Omega_{1}\deriv{q_{20}}{T}=\frac{q_{10}}{q_{3i}\left(q_{3i}^{2}-q_{10}^{2}-q_{20}^{2}\right)^{3/2}} \, ,
\end{equation}
\begin{equation}
q_{10}(0)=q_{1i} \, ,\quad q_{20}(0)=0 \, .
\end{equation}
Dividing Eq.~\eqref{eq:Dromo_eq_q10T} by Eq.~\eqref{eq:Dromo_eq_q20T} and integrating, a first integral is found in the form: 
\begin{equation}\label{eq:first_int_qx0}
q_{10}^{2}(T)+q_{20}^{2}(T)=q_{1i}^{2} \, .
\end{equation}
Introducing the first integral into {Eqs.~(\ref{eq:Dromo_eq_q10T},\ref{eq:Dromo_eq_q20T})} and further simplifying them with the choice of: 
\[
\Omega_{1}=\frac{1}{q_{3i}\left(q_{3i}^{2}-q_{1i}^{2}\right)^{3/2}},
\]
they can be solved to yield:
\begin{equation}
q_{10}(T)=q_{1i}\cos T\,,\quad q_{20}(T)=q_{1i}\sin T \, .\label{eq:Omega1_q10_q20}
\end{equation}
Because these values of $q_{10}$ and $q_{20}$ have been obtained by canceling the secular terms in Eq.~\eqref{eq:MTS_sol}, the expressions for $q_{11}$ and $q_{21}$ finally take the form:
\begin{equation}
\left[\begin{array}{c}
q_{11}\\
q_{21}
\end{array}\right]=\left[\begin{array}{c}
\mathcal{P}_{1}\\
\mathcal{P}_{2}
\end{array}\right] + \tan^{-1}\mathcal{K} \left[\begin{array}{c}
\mathcal{S}_{1}\\
\mathcal{S}_{2}
\end{array}\right]+\left[\begin{array}{c}
g_{1}\\
g_{2}
\end{array}\right].
\end{equation}

What remains to be determined for a complete solution of the first order terms $q_{11}$ and $q_{21}$ are the unknown functions $g_{1}$ and $g_{2}$. They can be obtained by looking at the order $\mathcal{O}(\varepsilon^{2})$ terms in the expansion of Eq.~\eqref{eq:Dromo_radial_ms_q1q2} from which one obtains the partial differential equation: 
\begin{equation}
\frac{\partial}{\partial\tau}\begin{bmatrix}q_{12}\\
q_{22}
\end{bmatrix}=-\Omega_{1}\frac{\partial}{\partial T}\left(\begin{bmatrix}q_{11}\\
q_{21}
\end{bmatrix}+\Omega_{2}\begin{bmatrix}q_{10}\\
q_{20}
\end{bmatrix}\right)+\frac{2s_{1}}{q_{3i}s_{0}^{3}}\begin{bmatrix}-\sin\tau\\
\cos\tau
\end{bmatrix}.\label{eq:Dromo_radial_ms_o2}
\end{equation}

The functions $g_{1}$ and $g_{2}$ as well as the coefficient $\Omega_{2}$ of the series expansion for $\Omega$ can, in principle, be determined by integrating the second member with respect to $\tau$ and imposing the cancellation of secular terms in $\tau$. Unfortunately the integration cannot be completed in closed analytical form. An approximate solution for the case of small initial eccentricity is however possible and is developed in Appendix~\ref{ap:g_functions} finally providing:
\[
g_{1}(T;q_{1i})= g_1^{(0)}(T) + q_{1i} g_1^{(1)}(T) + q_{1i}^2 g_1^{(2)}(T) + \ldots \, ,
\]
\[
g_{2}(T;q_{1i})= g_2^{(0)}(T) + q_{1i} g_2^{(1)}(T) + q_{1i}^2 g_2^{(2)}(T) + \ldots \, ,
\]
\[
\Omega_{2} = \Omega_2^{(0)} + q_{1i} \Omega_2^{(1)} + q_{1i}^2 \Omega_2^{(2)} + \ldots \, , 
\]
where:
\begin{equation*}
g_{1}^{(0)}(T)=\Omega_{1}^{3/4}\left(1+\cos T\right)\,,\quad g_{2}^{(0)}(T)=\Omega_{1}^{3/4}\sin T \, ,
\end{equation*}
\begin{equation*}
g_{1}^{(1)}(T)=0\,,\quad g_{2}^{(1)}(T)=-\Omega_{1}\sin T \, ,
\end{equation*}
\begin{equation*}
g_{1}^{(k)}(T)=\Omega_{1}^{\frac{3+k}{4}}\left[C_{0}^{(k)}+\sum_{l=1}^{k}C_{l}^{(k)}\cos lT\right] \quad k \geq 2 \, ,
\end{equation*}
\begin{equation*}
g_{2}^{(k)}(T)=-\deriv{g_{1}^{(k)}}{T}=\Omega_{1}^{\frac{3+k}{4}}\sum_{l=1}^{k}lC_{l}^{(k)}\sin lT \quad k \geq 2 \, ,
\end{equation*}
with the coefficients $\Omega_2^{(k)}$ and $C_l^{(k)}$ reported in Appendix~\ref{ap:g_functions}.

\subsection{Physical Time}

A multiple scales expression for the non-dimensional physical time can be obtained using the previous results and the time element presented in Eqs.~{(\ref{eq:Dromo_TE_t_relation},\ref{eq:Dromo_TE_t_ODE})}. Introducing the new independent variables $\tau$ and $T$ and expanding in power series of $\varepsilon$, the time element takes the form: 
\[
\zeta_0(\tau,T)=\zeta_{00}(\tau,T)+\varepsilon \zeta_{01}(\tau,T)+\mathcal{O}(\varepsilon^{2})\, .
\]
The zeroth order term $\zeta_{00}$ is given by the differential equation in $\tau$ which results from setting $\varepsilon=0$ in Eq. \eqref{eq:Dromo_TE_t_ODE}. After some manipulation, this yields:
\begin{equation}
\zeta_{00}(\tau,T)=q_{3i}\Omega_{1}\tau+g_t\left(T\right)\, , \label{eq:radial_ms_zeta00}
\end{equation}
where the additive function $g_t$ comes from imposing the secularity condition to the solution for $\zeta_{01}$. An approximation for small initial eccentricity is given in Appendix~\ref{ap:gtime_functions}, in the form of a power series expansion in $q_{1i}$.

The introduction of the time element allows us to simplify the multiple scales expression for the physical time by separating its secular and oscillatory components, with the former corresponding entirely to the time element. This structure is similar to the one obtained for the regular expansion, but now there is no spurious secular growth in the oscillatory components. Moreover, the slope of the secular part, which is a straight line in the fictitious time, differs from the one obtained for the regular expansion due to the contribution in the slow scale $T$ coming from $g_t$, which represents a small correction depending on the thrust parameter $\varepsilon$.

\subsection{Orbital Elements}

The multiple scales solution for the Dromo generalized orbital parameters presented so far can be used to calculate the classical orbital elements by direct application of {Eqs.~(\ref{eq:Dromo_fromDromo1},\ref{eq:Dromo_fromDromo2})}. Nevertheless, it is interesting to also express them as a power series expansion in $\varepsilon$, to better appreciate their physical behavior. Introducing the expressions for $q_{1}(\tau,T)$ and $q_{2}(\tau,T)$ into the first of Eqs.~\eqref{eq:Dromo_fromDromo1} and expanding in power series of $\varepsilon$ up to order one, the eccentricity can be given as follows: 
\begin{equation}
e(\tau,T;\varepsilon)=e_{0}+\frac{\varepsilon}{ q_{1i} q_{3i} } \left( q_{10} q_{11} + q_{20} q_{21} \right) +\mathcal{O}\left(\varepsilon^{2}\right) \, .
\end{equation}
For this expression to be valid, the term in $\varepsilon$ must be a small correction to the initial eccentricity $e_{0}$; otherwise, the first of Eqs.~\eqref{eq:Dromo_fromDromo1} must be used. In both cases, the effect on the eccentricity of the low radial thrust is a bounded variation around its initial value $e_{0}$, with an amplitude modulated by $\varepsilon$.

The non-dimensional orbital radius takes the form: 
\begin{equation}
r(\tau,T;\varepsilon)=\frac{1}{q_{3i}s_{0}}-\varepsilon\frac{s_{1}}{q_{3i}s_{0}^{2}}+\mathcal{O}(\varepsilon^{2}) \, .
\end{equation}
This expression for $r$ is bounded, as it should [see \cite{battin1999introduction}], and the leading order term is oscillatory with period $2\pi/(1-\Omega)$.

The Keplerian energy presents small oscillations about the constant zeroth order term: 
\begin{equation}
E(\tau,T;\varepsilon)= - \frac{q_{3i}^{2}-q_{1i}^{2}}{2}+\varepsilon \left( q_{10} q_{11} + q_{20} q_{21} \right) +\mathcal{O}\left(\varepsilon^{2}\right)\,,
\end{equation}
while the total energy takes the form: 
\[
E_{\text{T}}(\tau,T;\varepsilon)=-\frac{q_{3i}^{2}-q_{1i}^{2}}{2}+\varepsilon\left( q_{10} q_{11} + q_{20} q_{21} - \frac{1}{q_{3i} s_0} \right)+\mathcal{O}\left(\varepsilon^{2}\right) \, .
\]
The conservation of $E_{\text{T}}$ up to first order requires the term inside brackets to be constant. After some manipulations, this term can be written as:
\begin{equation*}
q_{10} g_{1}+q_{20}g_{2}-\frac{1}{q_{3i}\left(q_{3i}-q_{10} \right) } \, ,
\end{equation*}
which is a function of only the slow scale $T$. Imposing the conservation of $E_\text{T}$ in $T$ finally yields:
\begin{equation}
q_{10}g_{1}+q_{20}g_{2}=\frac{1}{q_{3i}\left(q_{3i}-q_{10} \right)}-\frac{1}{q_{3i}\left(q_{3i}+q_{1i}\right)} \, . \label{eq:firstint_g}
\end{equation}
By introducing the expressions for $g_{1}$ and $g_{2}$ given in Appendix~\ref{ap:g_functions} and expanding in power series of $q_{1i}$, it is checked that the conservation of energy is fulfilled for all the computed orders of $g_{1}$ and $g_{2}$, leaving a residual of $\mathcal{O}(q_{1i}^{l+1})$ where $l$ is the highest order of the expansions of $g_{1}$ and $g_{2}$. Moreover, this result can be used to improve the accuracy of the expressions for $e$ and $E$ by eliminating the direct dependence with $g_{1}$ and $g_{2}$: 
\begin{equation}
q_{10} q_{11} + q_{20} q_{21} = \frac{1}{q_{3i} s_0} -\frac{1}{q_{3i}q_{1i}\left(q_{3i}+q_{1i}\right)} \, .
\end{equation}

The angle $\gamma$ between the initial and osculating eccentricity vector can be given as:
\begin{equation}\label{eq:radial_ms_gamma}
\gamma(\tau,T;\varepsilon)=T+\varepsilon \left[ \frac{1}{q_{1i}^{2}}\left[q_{10}\left(\mathcal{P}_{2}+g_{2}\right)-q_{20}\left(\mathcal{P}_{1}+g_{1}\right)\right]+2\Omega_{1}\arctan\mathcal{K} \right] +\mathcal{O}(\varepsilon^{2}) .
\end{equation}
The zeroth order term, which was absent from the regular expansion, shows a secular behavior in the slow scale. This can be compared with the regular expansion having a secular component in $\theta$ inside its first order term. On the other hand, the first order term is now periodic and bounded. Note that this expansion is not valid for initially circular or quasi-circular orbits.

\subsection{Initially circular orbit}

The particular case of initially circular orbit is especially interesting for this multiple scales solution, since the expressions given for $g_{1}$ and $g_{2}$ are then exact. Particularizing the previous results for $e_{0}=0$ yields: 
\[
q_{1}(\tau,T;\varepsilon)=\frac{\varepsilon}{q_{3i}^{3}}\left(\cos\frac{T}{q_{3i}^{4}}-\cos\tau\right)+\mathcal{O}(\varepsilon^{2})\,,
\]
\[
q_{2}(\tau,T;\varepsilon)=\frac{\varepsilon}{q_{3i}^{3}}\left(\sin\frac{T}{q_{3i}^{4}}-\sin\tau\right)+\mathcal{O}(\varepsilon^{2})\,.
\]
Although $q_{10}$ and $q_{20}$ no longer appear, the variation with the slow scale is retained through the contributions from $g_{1}$ and $g_{2}$. Comparing these expressions with those obtained for the regular expansion, it is straightforward to check that the latter coincide with the former for $T=0$.

From this solution, it is possible to derive: 
\[
e(\tau,T;\varepsilon)=\varepsilon\frac{\sqrt{2}}{q_{3i}^{4}}\left[1-\cos\left(\tau-\frac{T}{q_{3i}^{4}}\right)\right]^{1/2}+\mathcal{O}(\varepsilon^{2}) \, ,
\]
\[
r(\tau,T;\varepsilon)=\frac{1}{q_{3i}^{2}}+\varepsilon\frac{1}{q_{3i}^{6}}\left[1-\cos\left(\tau-\frac{T}{q_{3i}^{4}}\right)\right]+\mathcal{O}(\varepsilon^{2}) \, ,
\]
\[
E\left(\tau,T;\varepsilon\right)=-\frac{q_{3i}^{2}}{2}+\mathcal{O}(\varepsilon^{2}) \, ,
\]
\[
E_{\text{T}}\left(\tau,T;\varepsilon\right)=-\frac{q_{3i}^{2}}{2}-\varepsilon\frac{1}{q_{3i}^{2}}+\mathcal{O}(\varepsilon^{2}) \, ,
\]
\[
\gamma(\tau,T;\varepsilon)=\arctan\frac{\sin(T/q_{3i}^{4})-\sin\tau}{\cos(T/q_{3i}^{4})-\cos\tau}+\mathcal{O}(\varepsilon) \, .
\]
Note that the leading order terms of $e$ and $r$ now show a new period $2\pi/\left(1-\Omega/q_{3i}^{4}\right)$ in $\theta$, resulting from the combination of the periods for the fast and slow `time' scales.

\subsection{Eccentricity vector rotational frequency}\label{sec:eccentricity_vactor_rotational_frequency}

\begin{figure}
\centering \includegraphics[width=0.6\columnwidth]{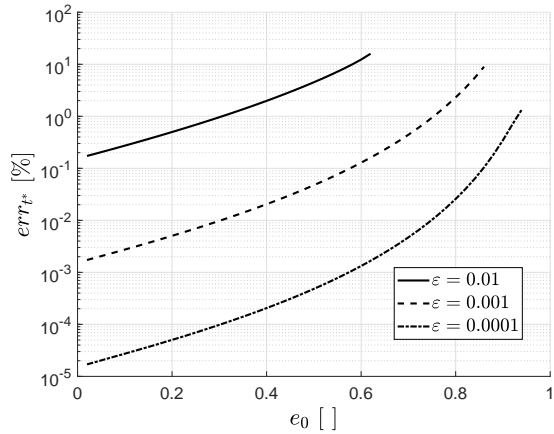}
\caption{Percentage error in the time estimation for the eccentricity
vector to complete a whole revolution, \mbox{$err_{t^{*}}=|t^{\text{ms}}/t^{\text{num}}-1|/100$}, where $t^{\text{ms}}$ is the multiple scales solution and $t^{\text{num}}$ is a high-precision numerical solution. The curves are obtained up until escape conditions are reached.}\label{fig:err_t_eccvectrev} 
\end{figure}

The number of orbits it takes for the eccentricity vector to describe a whole turn can be approximated through the leading order term in Eq.~\eqref{eq:radial_ms_gamma}, yielding: 
\[
\frac{\theta^{*}}{2\pi}=\frac{1}{\Omega} \, .
\]
It is interesting to express this result in terms of the physical time $t$. From Eq.~\eqref{eq:Dromo_TE_t_relation} it follows that the only contributions to the secular evolution of $t$ are those coming from the time element $\zeta_0$; then, using Eq.~\eqref{eq:radial_ms_zeta00} it is possible to write: 
\begin{equation}
t^{*}=\left(q_{3i}\Omega_{1}+D_t\Omega\right)\theta^{*}=2\pi\left(\frac{q_{3i}}{\varepsilon\left(1+\Omega_{2}\varepsilon\right)}+D_t\right) \, ,
\end{equation}
with the same $D_t$ reported in Appendix~\ref{ap:gtime_functions} for the solution of $g_t$. Figure~\ref{fig:err_t_eccvectrev} shows the error committed by the previous expression compared with a high-precision numerical propagation. As expected, the accuracy of the approximation degrades as $\varepsilon$ and $e_{0}$ increase.

\subsection{Validity of the expansion}\label{sec:validity_of_the_expansion}

\begin{figure}
\centering \includegraphics[width=0.6\columnwidth]{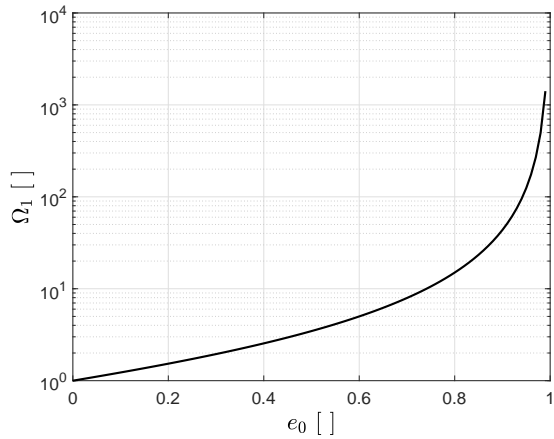}
\caption{Evolution of parameter $\Omega_{1}$ with the initial eccentricity $e_{0}$. Starting from $1$ for initially circular orbit, it grows rapidly with the ellipticity of the initial orbit.}\label{fig:Omega1_vs_e0} 
\end{figure}

\begin{figure}
\centering \includegraphics[width=0.6\columnwidth]{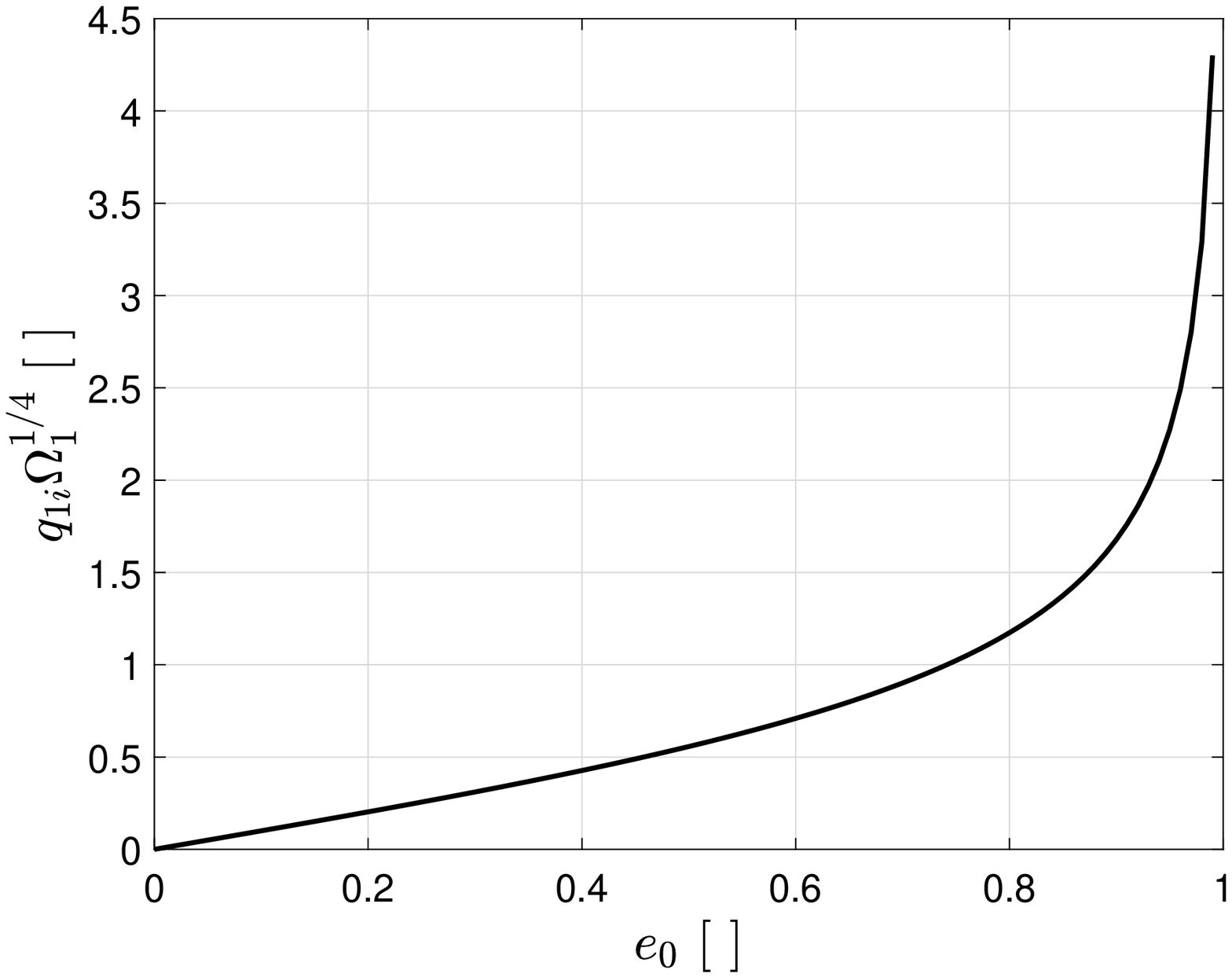}
\caption{Evolution of coefficient $q_{1i}\Omega_{1}^{1/4}$ with the initial eccentricity $e_{0}$. Being present in the power series expansions of $g_{1}$, $g_{2}$ and $\Omega_{2}$, its value has a direct impact on the validity of the asymptotic solution}\label{fig:q1iOmega1p1_4_vs} 
\end{figure}

The validity of the multiple scales solution depends on both the non-dimensional thrust parameter $\varepsilon$ and the initial eccentricity $e_{0}$. The influence of the former is easy to evaluate, whereas the latter requires a more careful study. Two different aspects have to be considered, namely the expanding interval of the fictitious time in which the results are valid and the asymptoticness of the additional expansions for $g_{1}$, $g_{2}$ and $\Omega_{2}$.

Before going into deeper detail, it is convenient to take a look at the evolution of $\Omega_{1}$ with $e_{0}$. As the ellipticity of the initial orbit is increased, the values of $q_{3i}$ and $q_{1i}$ approach each other and $\Omega_{1}$ grows rapidly; this behavior can be seen in Fig.~\ref{fig:Omega1_vs_e0}. It is also important to study the evolution of the factor $q_{1i}\Omega_{1}^{1/4}$, since it plays an important part in the expansions for $g_{1}$, $g_{2}$ and $\Omega_{2}$. As shown in Fig.~\ref{fig:q1iOmega1p1_4_vs}, it also grows with the initial eccentricity, although the values reached are noticeably smaller.

According to the theoretical basis of the method of multiple scales, and neglecting for the time being the influence of the extra expansion performed in $q_{1i}$, the results should be valid for an expanding interval of the fictitious time of length at least $\mathcal{O}(1/\Omega_{1}\varepsilon)$. Since a higher order formulation (i.e. including a coordinate strain for the slow `time' scale) is used a longer interval of $\mathcal{O}(1/\left(\Omega_{1}\varepsilon\right)^{2})$
could be expected, but there are no mathematical warranties for this [see \cite{murdock1999perturbations,bender2013advanced}]. Comparing the numerical results given in Section~\ref{sec:numerical_evaluation} with those obtained by \cite{Gonzalo2014} using a first order multiple scales formulation, a clear improvement in the expanding interval is observed for the higher order solution. In both cases the expanding interval length depends on $\varepsilon\Omega_{1}$, placing a limit to the values of $\theta$ for which the expansion remains valid as a function of $\varepsilon$ and $e_{0}$.

The additional regular expansion performed in $q_{1i}$ for the calculation of $g_{1}$, $g_{2}$ and $\Omega_{2}$ introduces further considerations regarding the validity of the solution with the initial eccentricity $e_{0}$. By careful examination of the results in Appendix~\ref{ap:g_functions}, it is checked that the $k\text{-th}$ term of the expansion for each of these functions is multiplied by:
\[
\Omega_{1}\left(q_{1i}\Omega_{1}^{1/4}\right)^{k}\,.
\]
Consequently, the behavior of these expansions will depend on the coefficient $q_{1i}\Omega_{1}^{1/4}$, whose evolution was represented in Fig.~\ref{fig:q1iOmega1p1_4_vs}. Considering only the contribution of this coefficient the validity of the expansions should break for values of $e_{0}$ around $0.74$, for which it becomes greater than one. However, the expansions also contain numerical coefficients that decrease as the order increases, offering an appreciable improvement. In any case, it is verified that the quality of the solution degrades rapidly for nearly parabolic initial orbits.

\subsection{Effect of mass variation for the constant-thrust case}
The multiple scales solution is obtained under the assumption of constant radial perturbing acceleration, leading to a fixed value of the non-dimensional thrust parameter $\varepsilon$. From a practical point of view, this can be the case for propellantless propulsion systems such as sails, or when the thrust magnitude $F$ varies in time to accommodate the reduction in mass. For a thruster with constant thrust magnitude, however, $\varepsilon$ will increase as the propellant mass is consumed. Nevertheless, because low-thrust thrusters (corresponding to continuous operation with $\varepsilon \ll 1$ as considered in this work) have a very small propellant mass flow rate, the assumption of constant $\varepsilon$ still holds. The long-term mass variation can then be included by following a rectification procedure like the one proposed by \cite{niccolai2018orbital}, which is an extended application of the one by \cite{bombardelli2011asymptotic}. While \cite{bombardelli2011asymptotic} only consider the rectification procedure to improve the accuracy of their solution by limiting the accumulation of errors in $\theta$, \cite{niccolai2018orbital} extend its use to account for the variations in $\varepsilon$. The application of the rectification procedure from these works to our multiple scales solution is straightforward, as they also deal with asymptotic solutions for constant acceleration problems using Dromo formulation (\cite{bombardelli2011asymptotic} tackles the tangential thrust case, whereas \cite{niccolai2018orbital} deals with the circumferential one).

From the physical point of view, the structure of the multiple scales solution reveals that the period of the fast time scale will remain unaffected by the change in mass, whereas the period of the slow one will decrease together with mass. This will, in turns, reduce the rotation period of the eccentricity vector introduced in Section~\ref{sec:eccentricity_vactor_rotational_frequency}.

\section{Numerical Evaluation of the Results}\label{sec:numerical_evaluation}

\begin{figure*}
  \centering
  \includegraphics[width=0.48\textwidth]{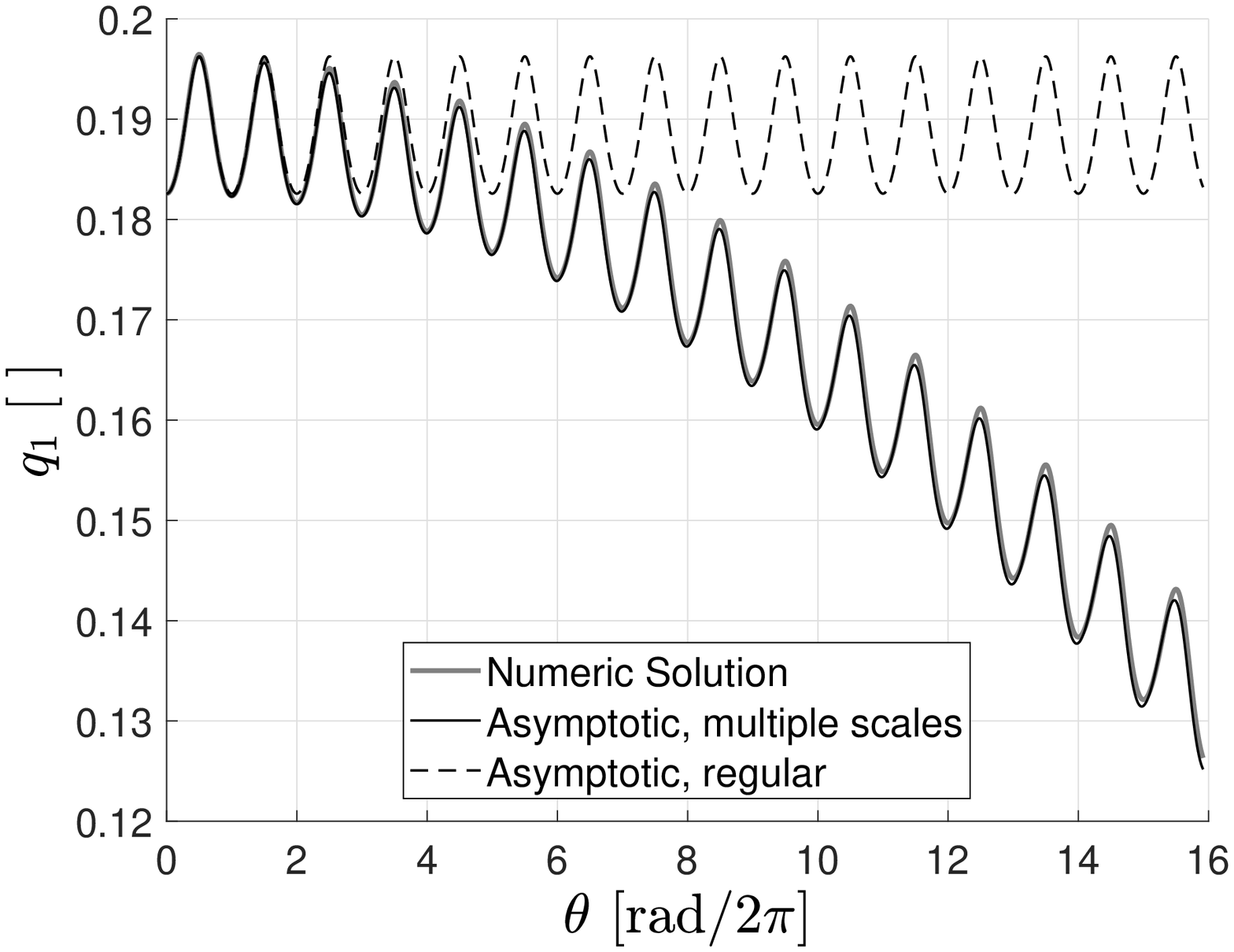}\hfill
  \includegraphics[width=0.48\textwidth]{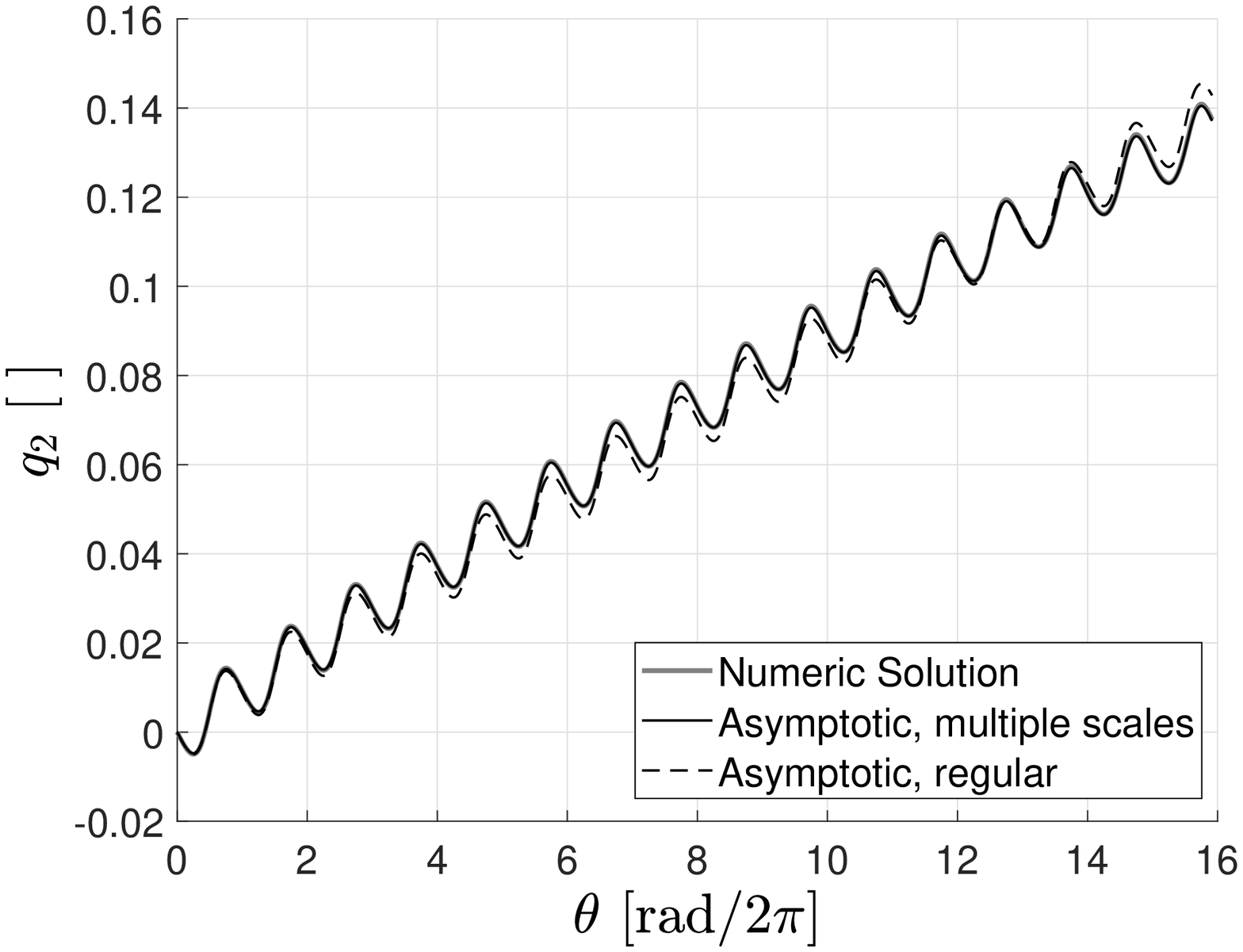} \\[1em]
  \includegraphics[width=0.48\textwidth]{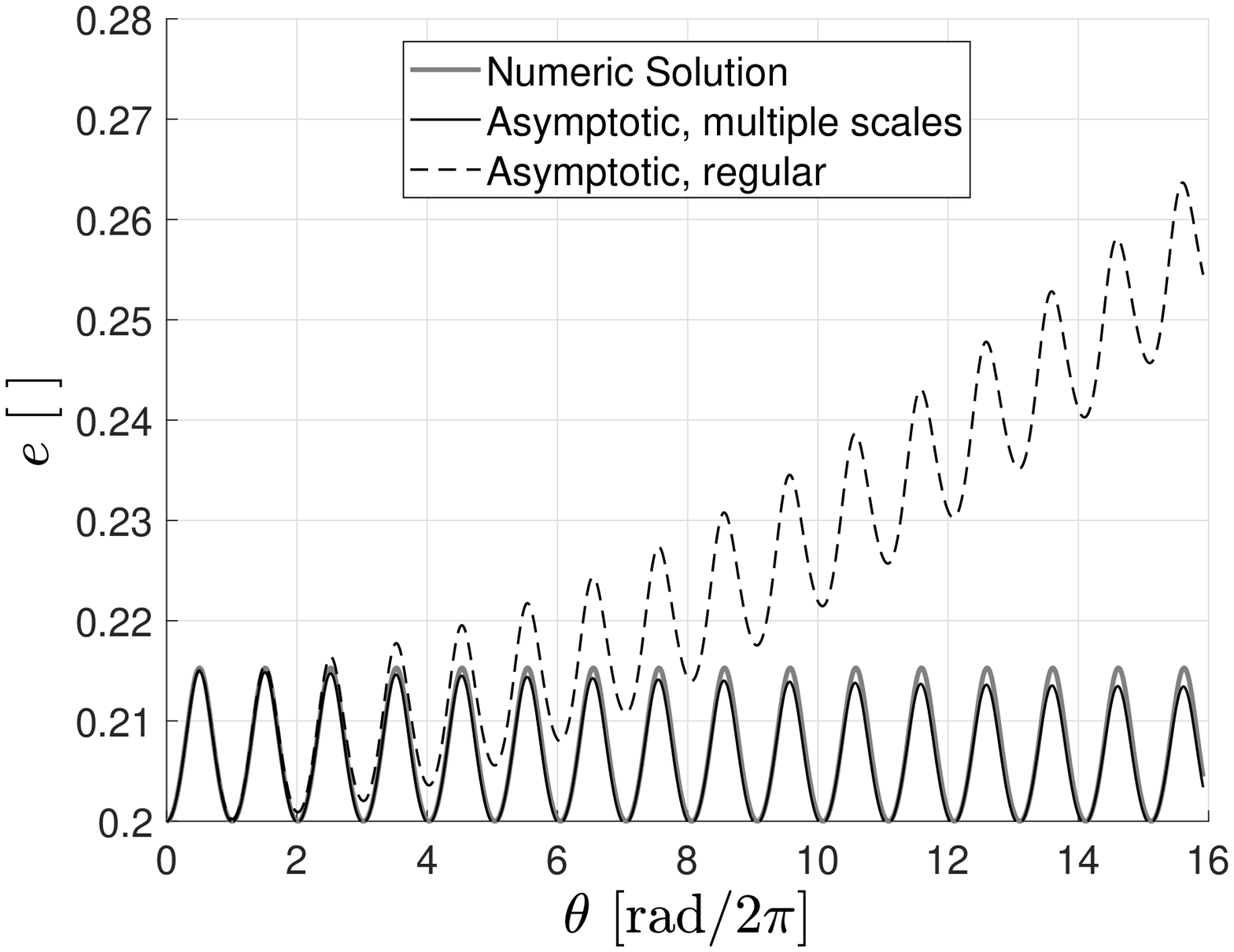}\hfill
  \includegraphics[width=0.48\textwidth]{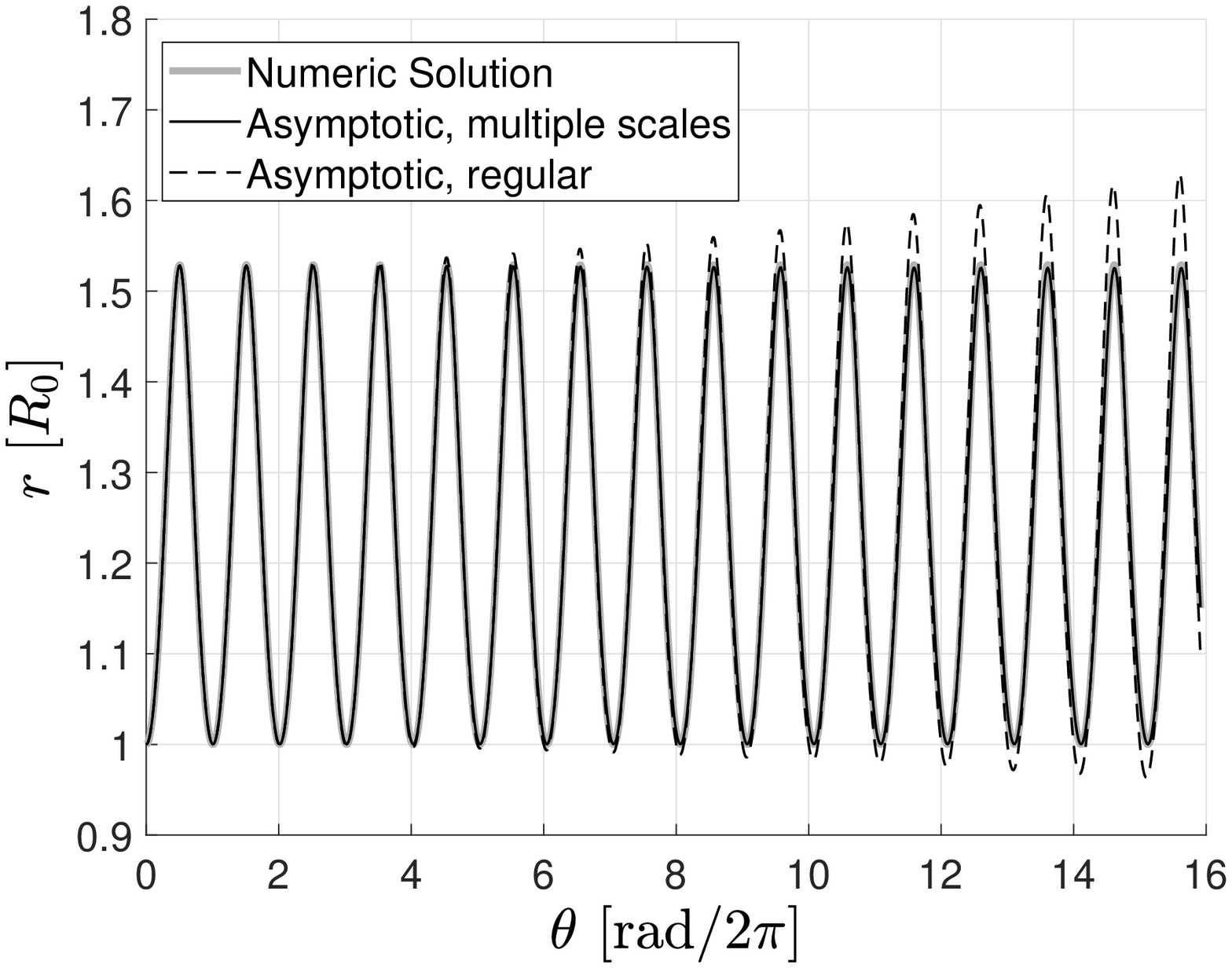} \\[1em]
  \includegraphics[width=0.48\textwidth]{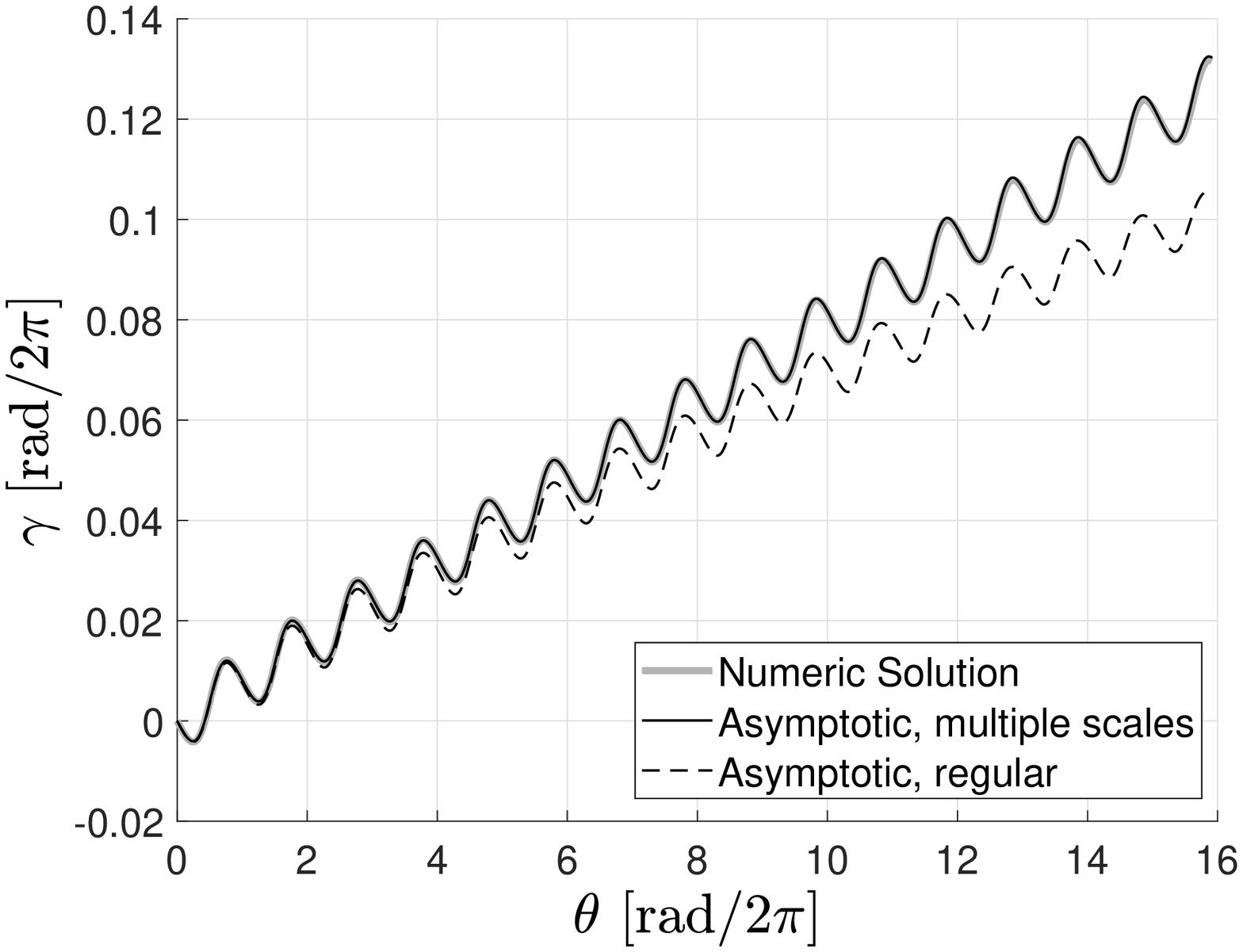}\hfill
  \includegraphics[width=0.48\textwidth]{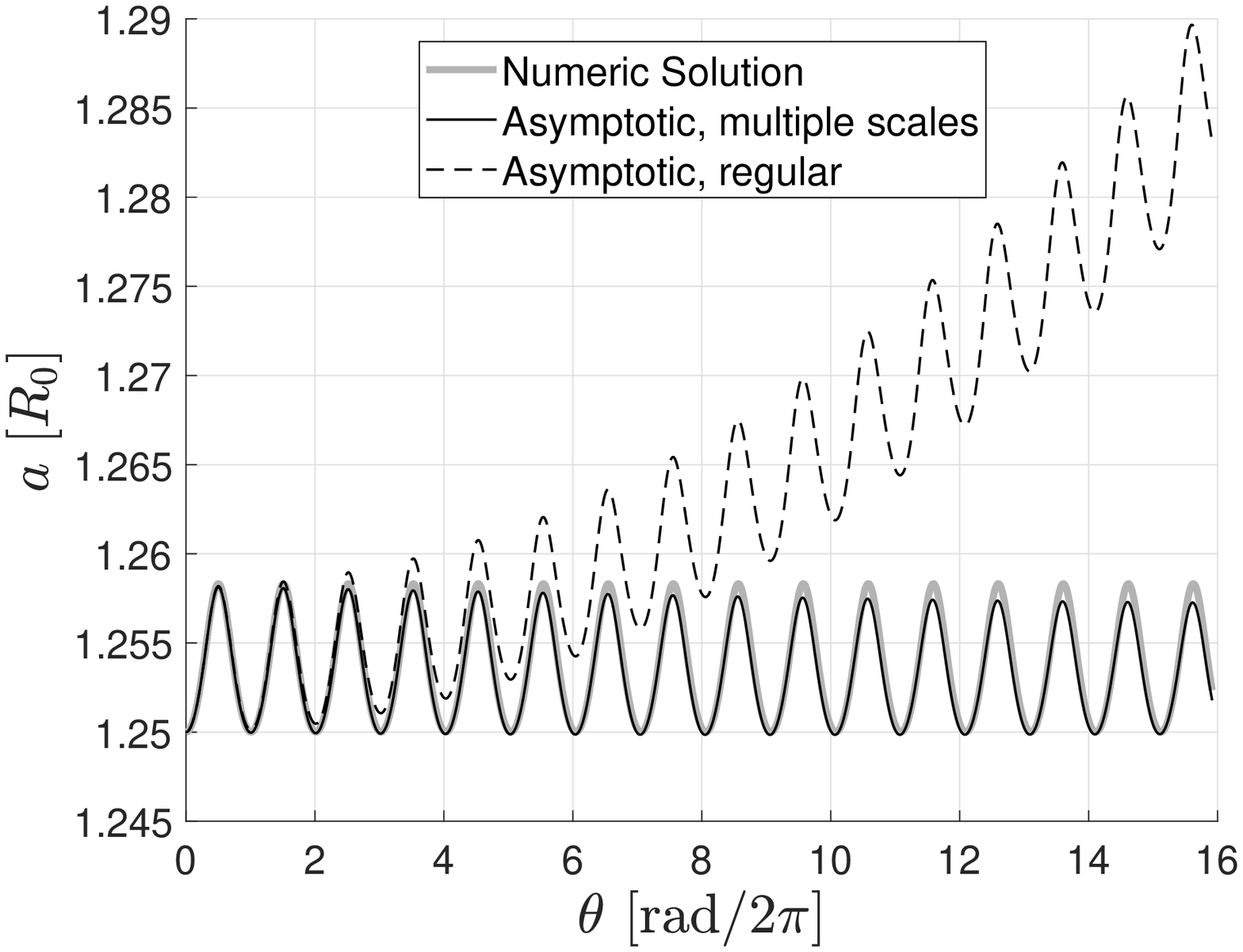}
  \caption{Comparison of the solution obtained for $q_{1}$, $q_{2}$, $e$, $r$, $\gamma$ and $a$ in the radial thrust case, for $e_{0}=0.2$ and $\varepsilon=0.005$.}\label{fig:radial_e02_eps0005} 
\end{figure*}

\begin{figure*}
  \centering
  \includegraphics[width=0.48\textwidth]{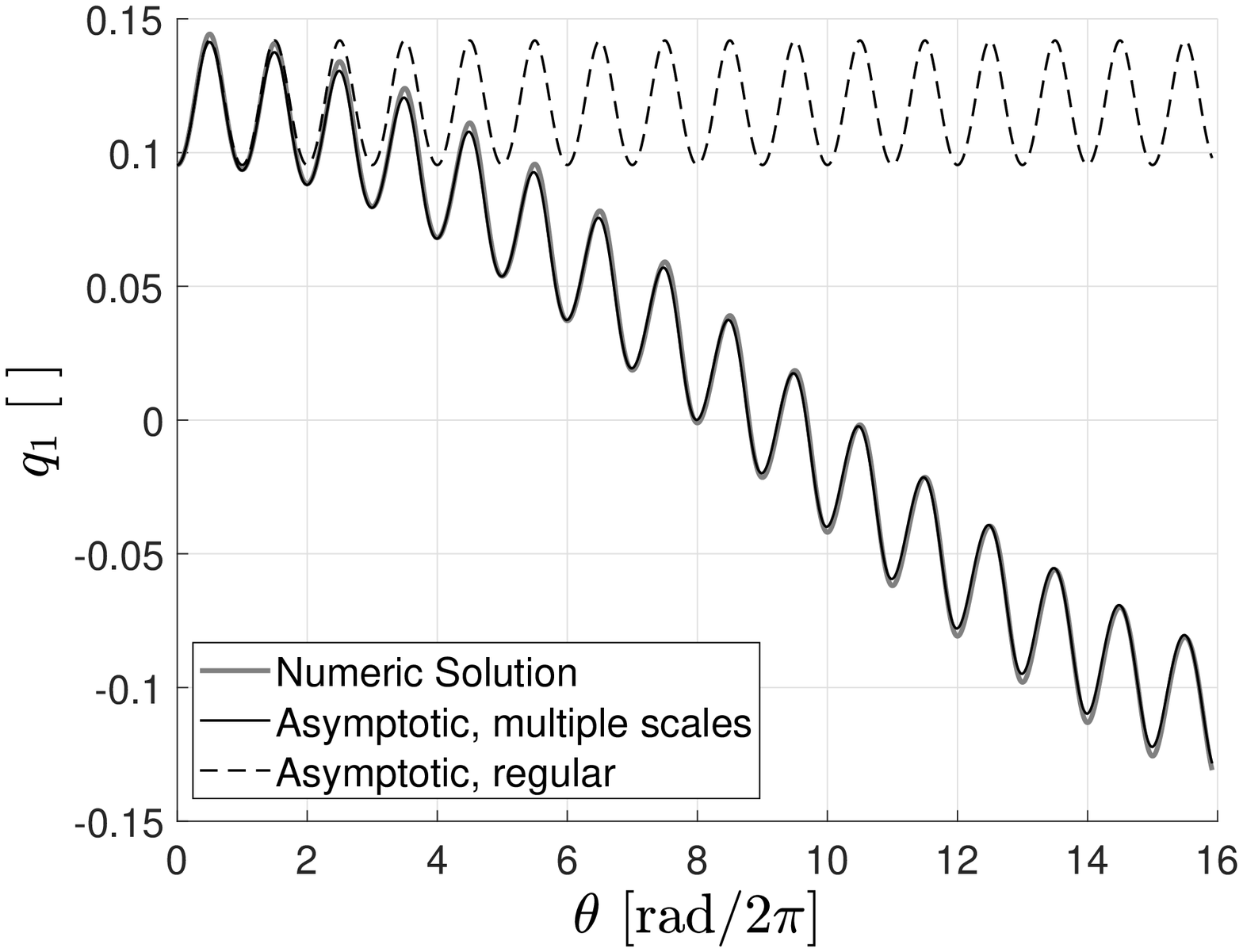}\hfill
  \includegraphics[width=0.48\textwidth]{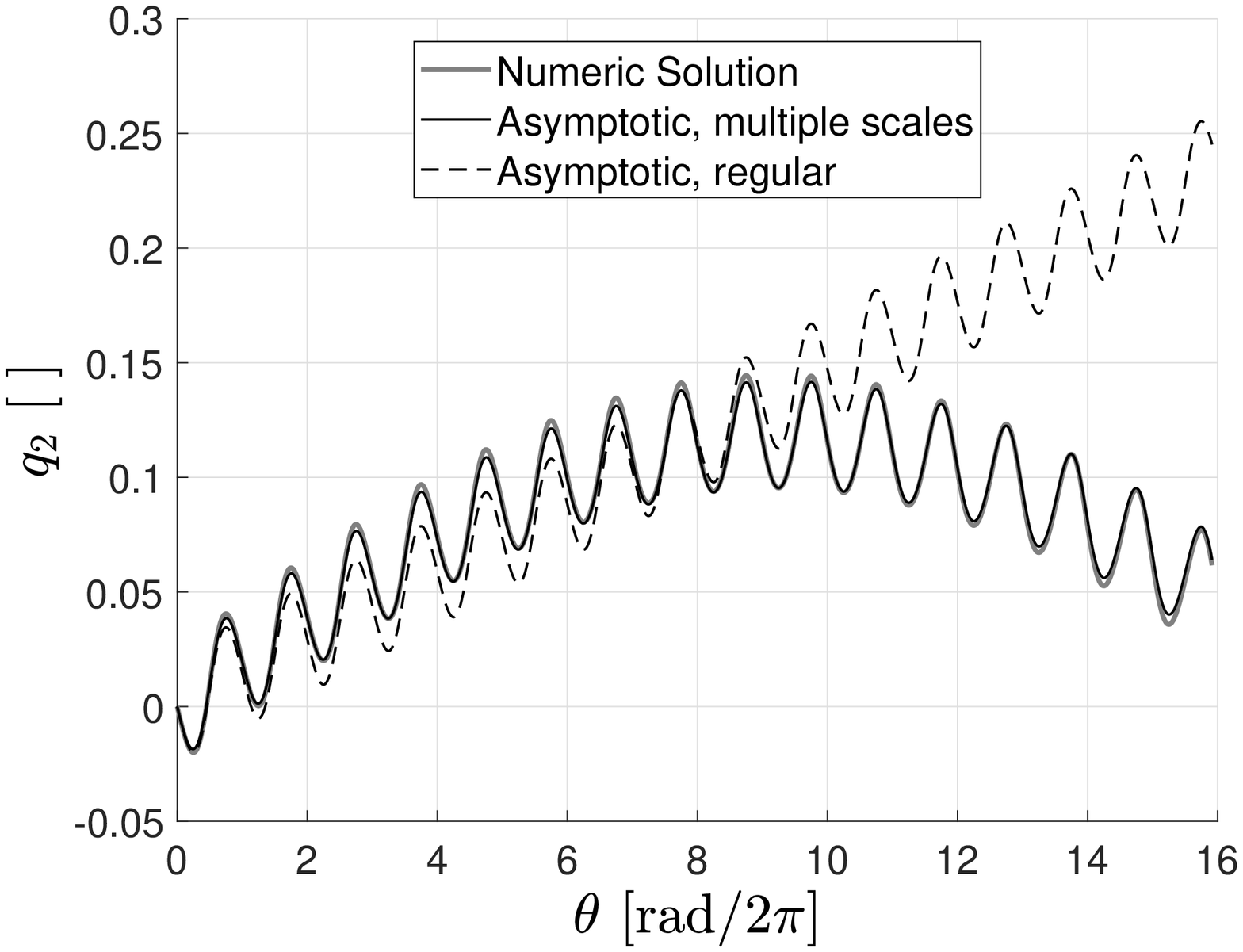} \\[1em]
  \includegraphics[width=0.48\textwidth]{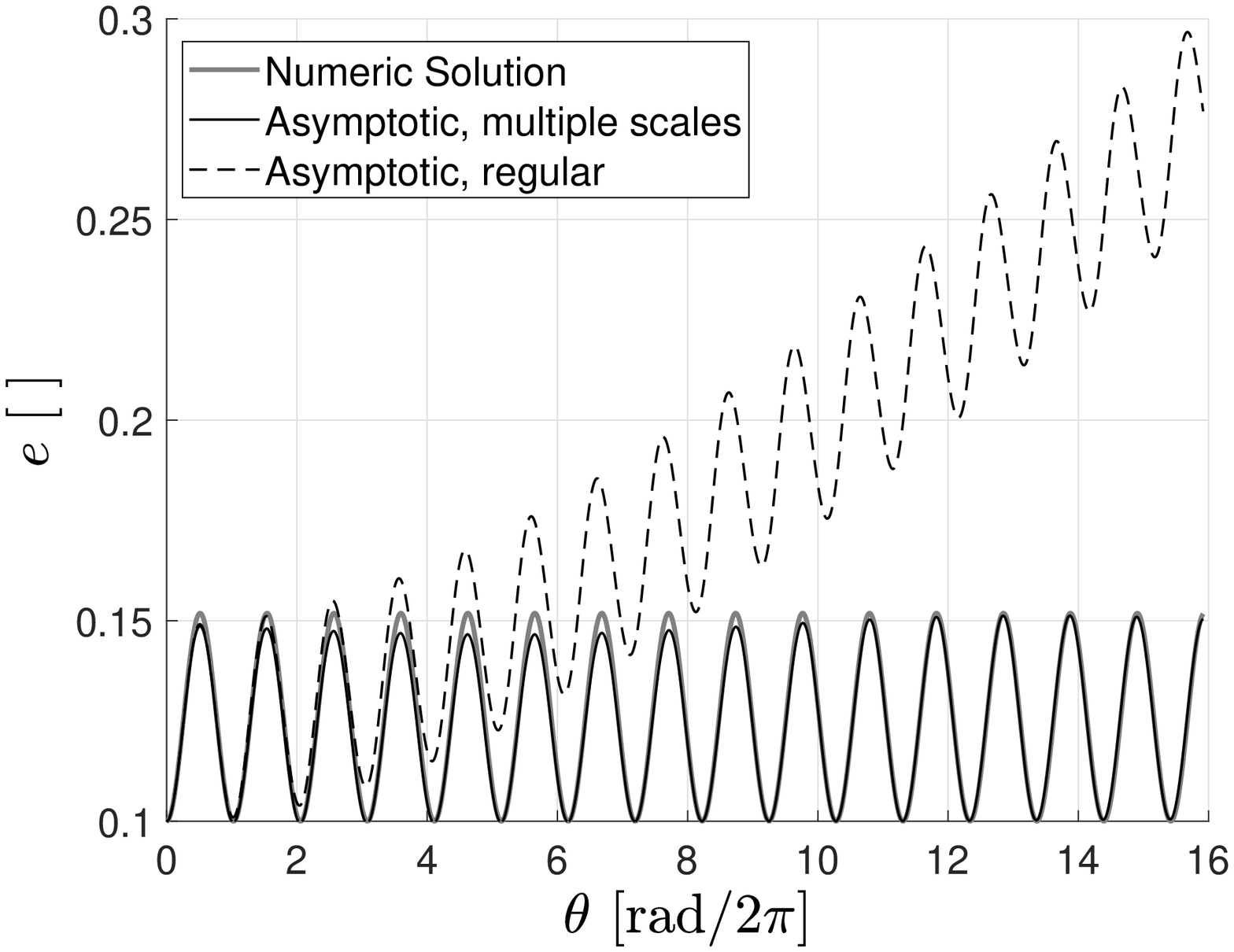}\hfill
  \includegraphics[width=0.48\textwidth]{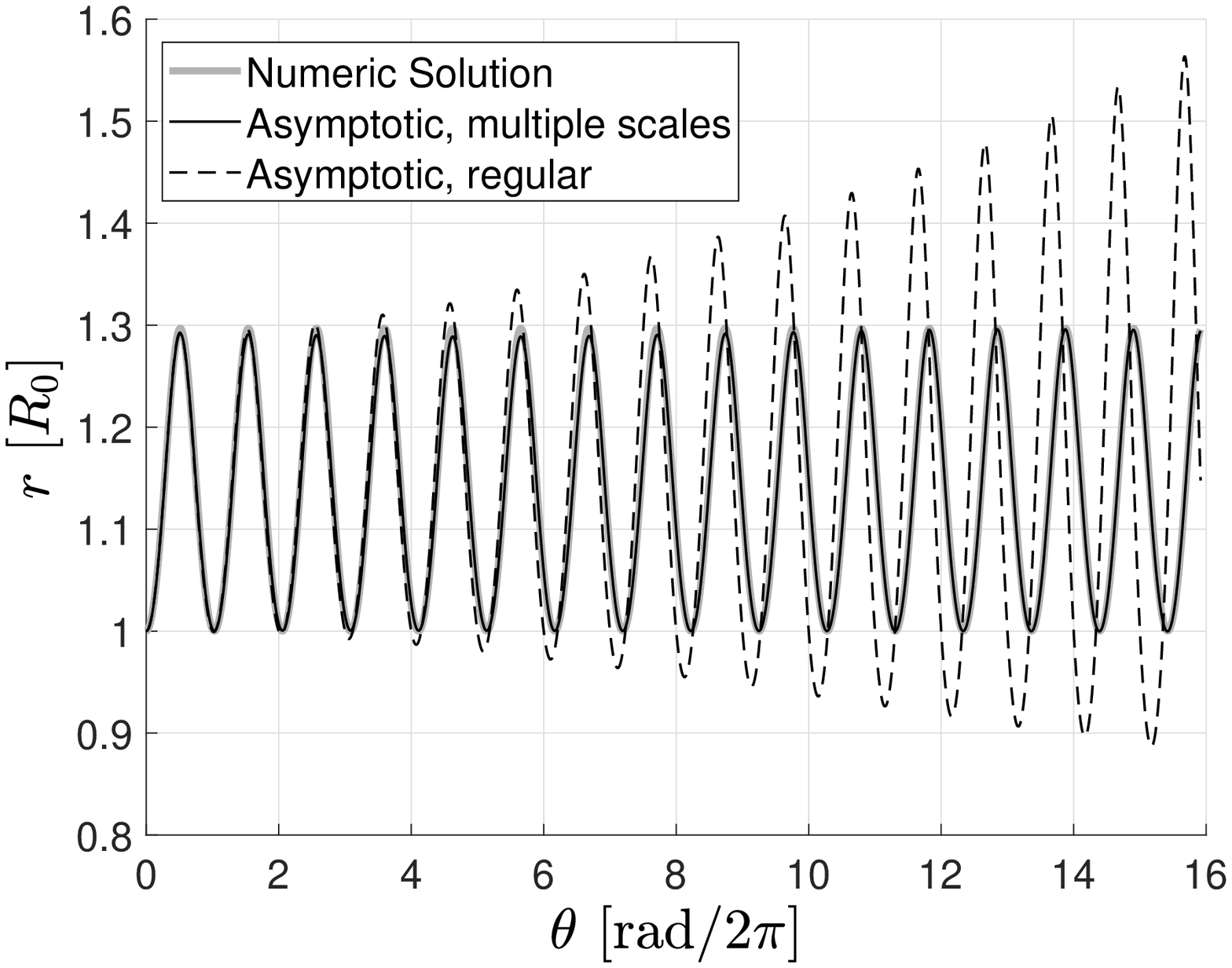} \\[1em]
  \includegraphics[width=0.48\textwidth]{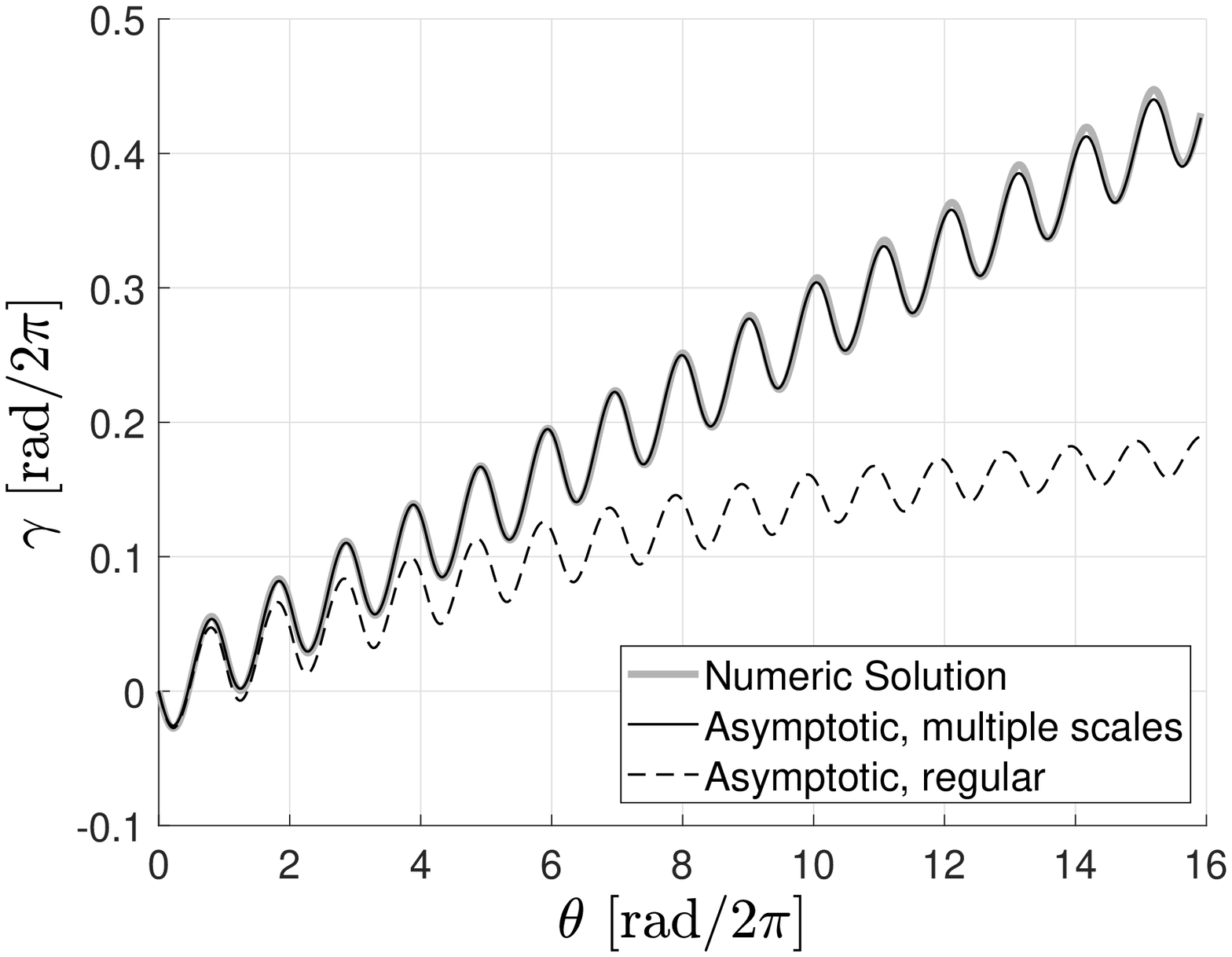}\hfill
  \includegraphics[width=0.48\textwidth]{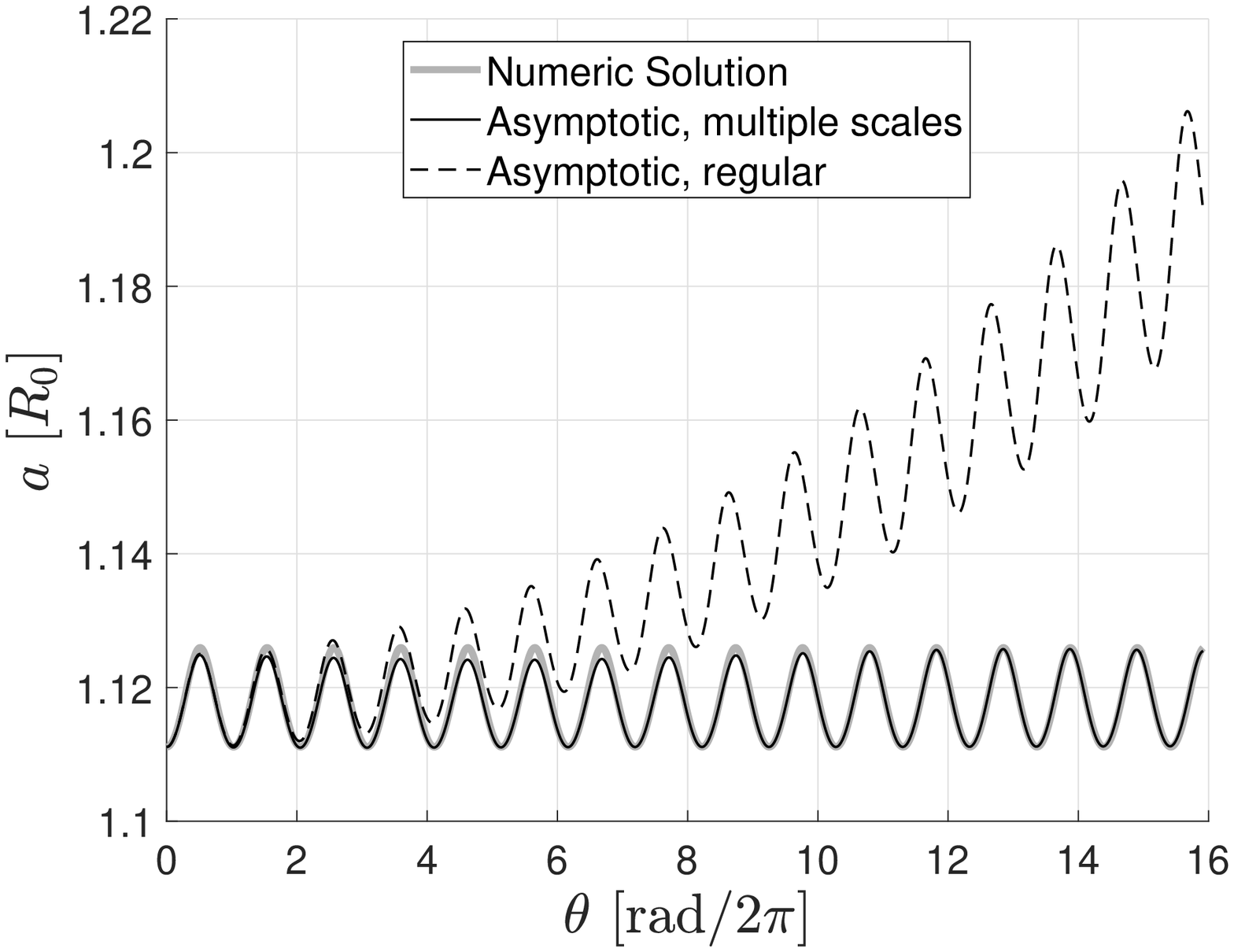}
  \caption{Comparison of the solution obtained for $q_{1}$, $q_{2}$, $e$, $r$, $\gamma$ and $a$ in the radial thrust case, for $e_{0}=0.1$ and $\varepsilon=0.02$.}\label{fig:radial_e01_eps002} 
\end{figure*}

\begin{figure*}
  \centering
  \includegraphics[width=0.48\textwidth]{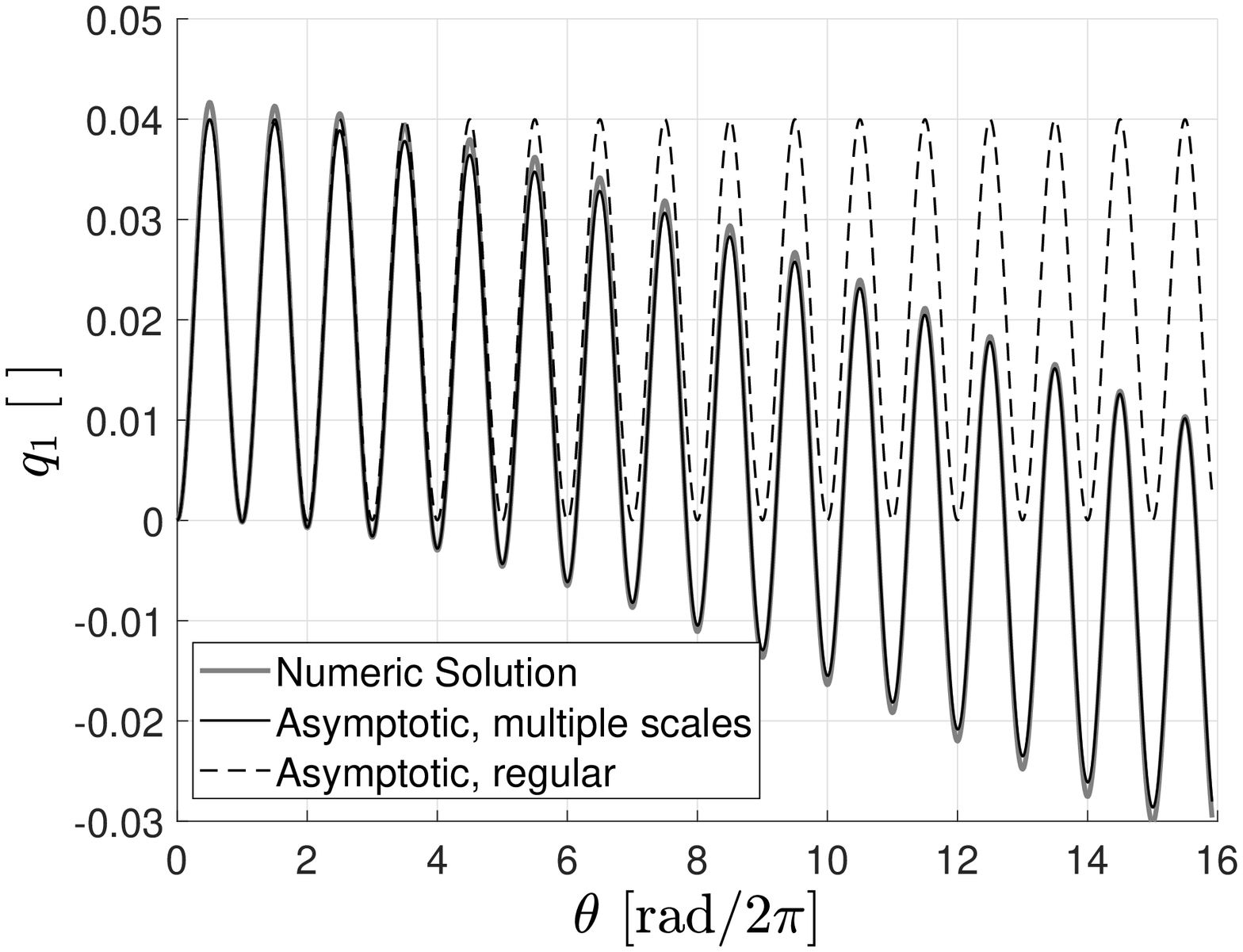}\hfill
  \includegraphics[width=0.48\textwidth]{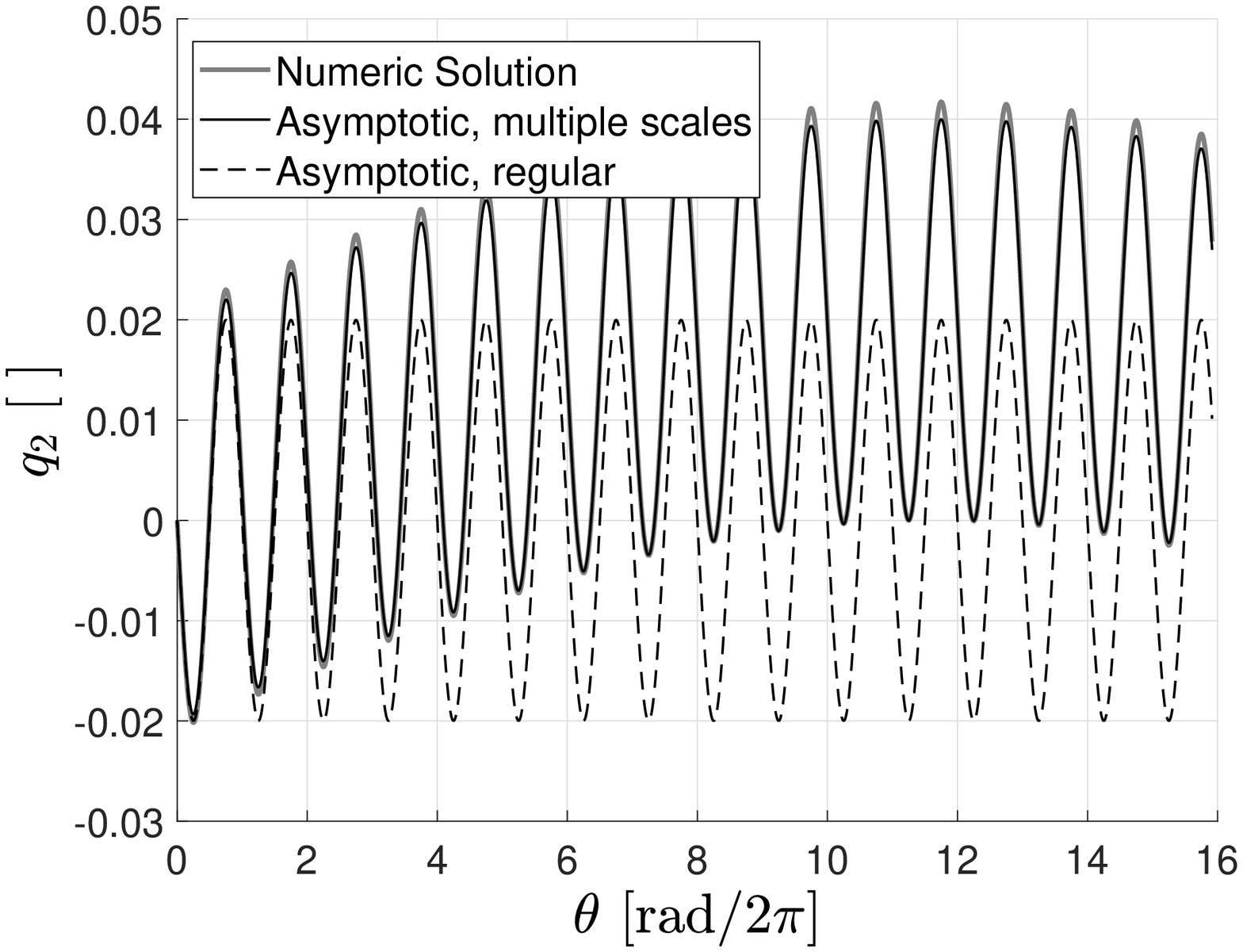} \\[1em]
  \includegraphics[width=0.48\textwidth]{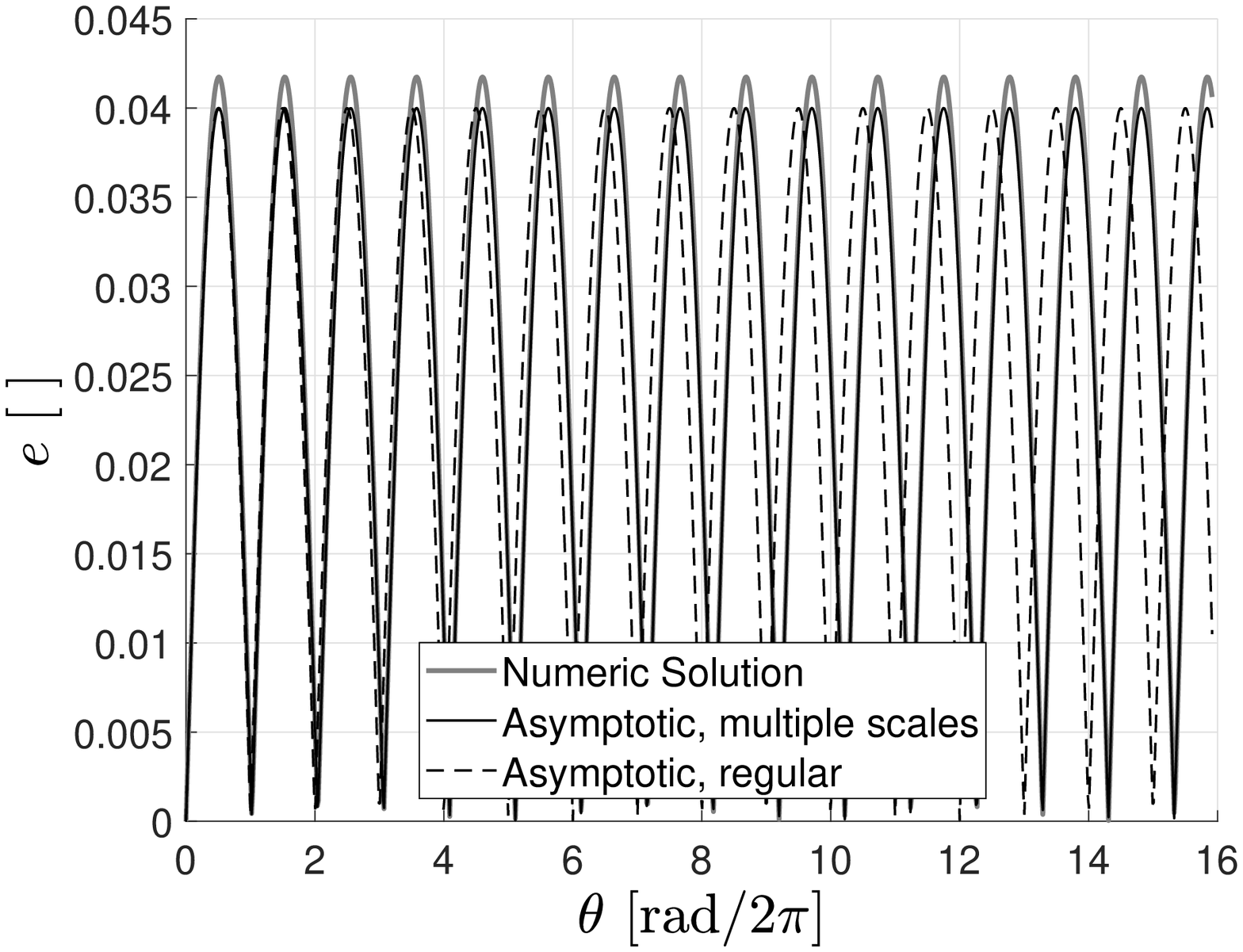}\hfill
  \includegraphics[width=0.48\textwidth]{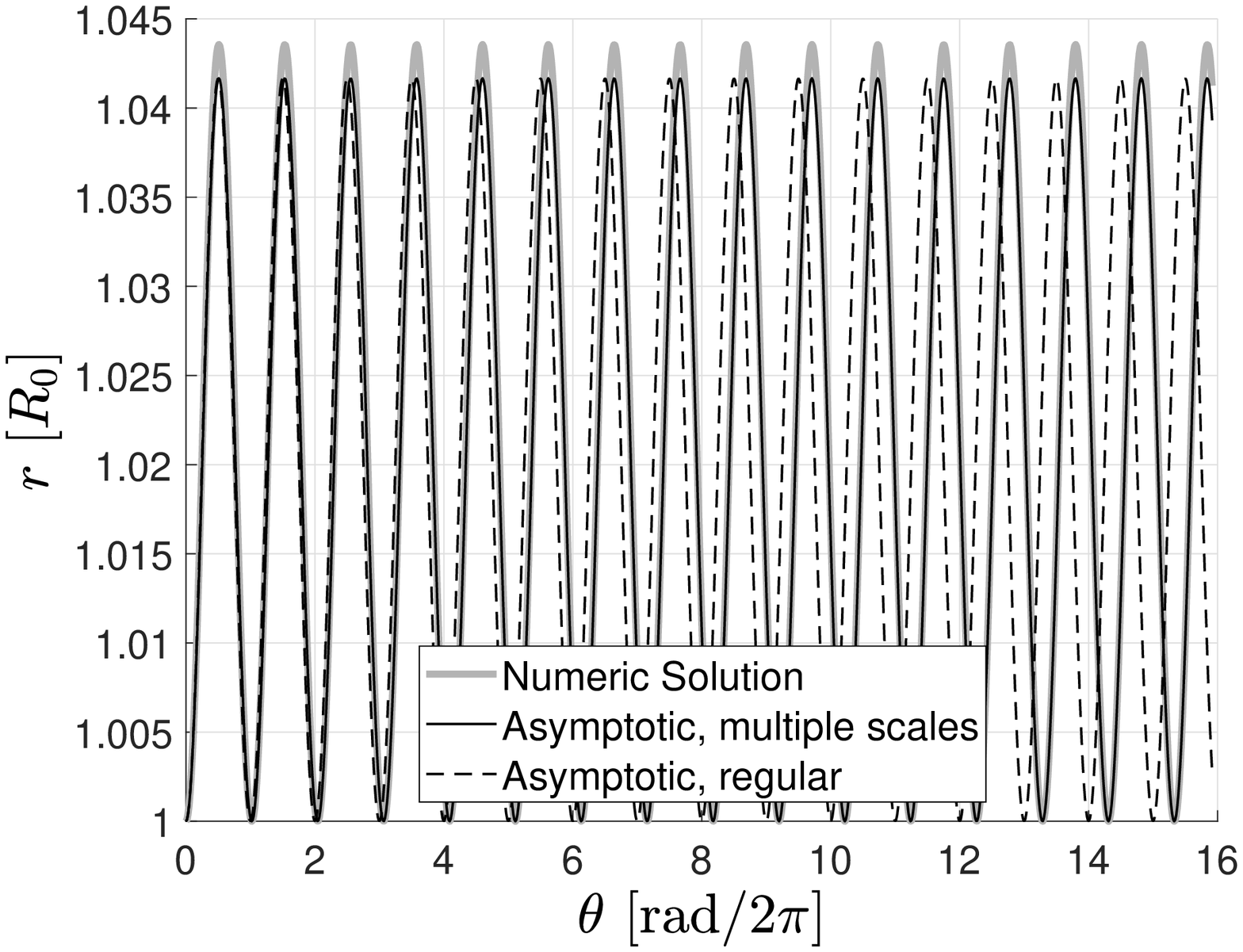} \\[1em]
  \includegraphics[width=0.48\textwidth]{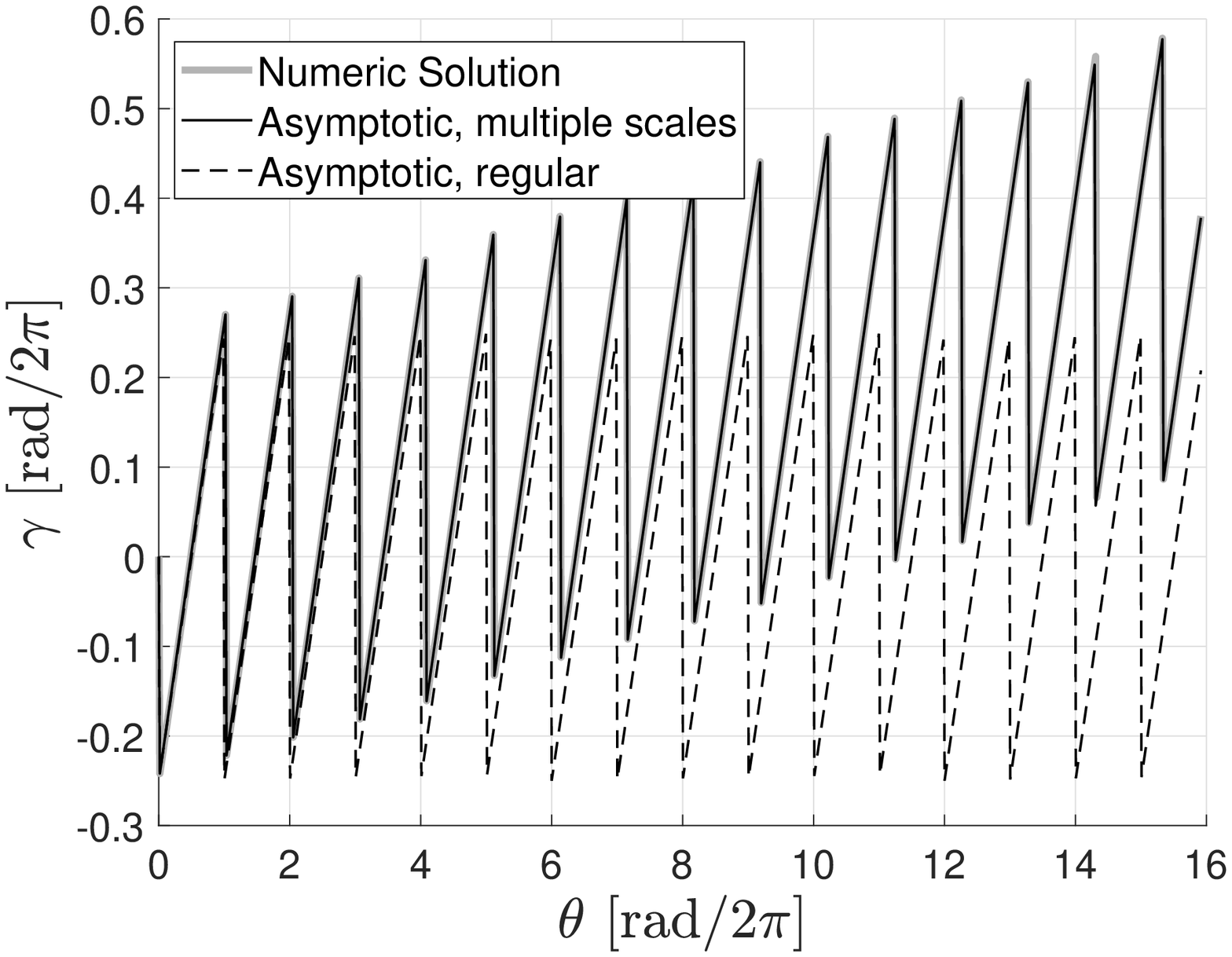}\hfill
  \includegraphics[width=0.48\textwidth]{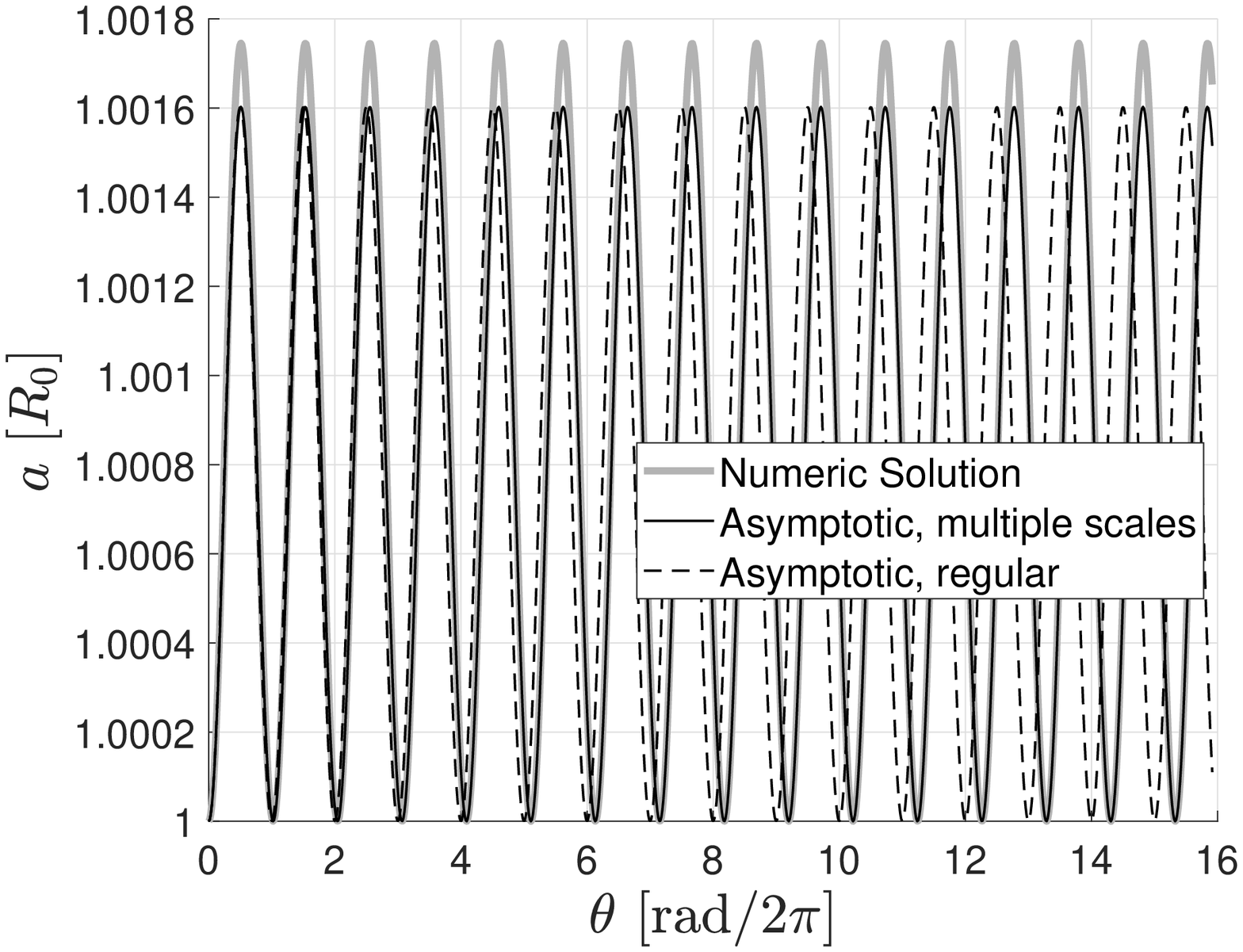}
  \caption{Comparison of the solution obtained for $q_{1}$, $q_{2}$, $e$, $r$, $\gamma$ and $a$ in the radial thrust case, for $e_{0}=0$ and $\varepsilon=0.02$.}\label{fig:radial_e0_eps002} 
\end{figure*}

In this section, the quality of the asymptotic solutions obtained so far is evaluated by comparing them with reference numerical solutions computed using a high-precision integrator. It will be seen that the use of multiple scales not only improves the quality of the results, but also provides interesting information about the physics of the problem.

Figures \ref{fig:radial_e02_eps0005}-\ref{fig:radial_e0_eps002} show the evolution of Dromo parameters $(q_{1},q_{2})$, eccentricity $e$, non-dimensional orbital radius $r$, angular displacement of the eccentricity vector $\gamma$ and non-dimensional semi-major axis $a$ for several values of the initial eccentricity $e_{0}$ and the non-dimensional thrust acceleration parameter $\varepsilon$. The values of $e$, $r$, $\gamma$ and $a$ have been calculated using Eqs.~(\ref{eq:Dromo_fromDromo1}-\ref{eq:Dromo_fromDromo2}) and Eq.~\eqref{eq:Dromo_r}. The first set of results, Fig.~\ref{fig:radial_e02_eps0005}, corresponds to the case of $\varepsilon=0.005$ and $e_{0}=0.2$; using Eq.~\eqref{eq:radial_Dromoic}, the initial values of the Dromo variables associated to this $e_{0}$ are $q_{1i}=0.1826$ and $q_{3i}=0.9129$. The first conclusion is that the regular expansion fails very soon for $q_{1}(\theta)$; the mean value remains constant, so it cannot reproduce the evolution in the slow scale. Its behavior is better for $q_{2}(\theta)$, since it contains a secular term that approximately reproduces the sinusoidal slow scale evolution for small values of $\theta$. It is important to highlight that the secular components in the regular solution correspond to the first terms of the Taylor expansions of $q_{10}(T)$ and $q_{20}(T)$. The multiple scales solution turns out to be remarkably good, slowly separating from the real one as $\theta$ grows. The results for $e$, $a$ and $r$ inherit the properties from $q_{1}(\theta)$ and $q_{2}(\theta)$; since the reference solutions for $e$, $a$ and $r$ oscillate between fixed values, the evolution in the slow scale of $q_{1}(\theta)$ and $q_{2}(\theta)$ must compensate each other to obtain a good approximation. This is not possible for the regular expansion, which only contains a secular term in $q_{2}(\theta)$. As a consequence, a spurious secular evolution appears for $e$ and $a$ separating them from the reference solution very soon, while the amplitude of $r$ increases with $\theta$ instead of remaining constant. On the other hand, the multiple scales formulation faithfully represents the real solution for the range of $\theta$ shown in the figures. Finally, a good agreement is observed between the reference and multiple scales solutions for $\gamma$, while the regular expansion slowly diverges from them.

Figure~\ref{fig:radial_e01_eps002} corresponds to an orbital propagation with $\varepsilon=0.02$ and $e_{0}=0.1$ ($q_{1i}=0.0935$, $q_{3i}=0.9535$). Most of the comments made for the previous case still hold, only now the separation between the reference and the multiple scales solutions grows faster with $\theta$. It is checked that the greater errors for the multiple scales solution come from the evolution of the mean values in the slow scale, not the amplitude or period of the oscillations in the fast scale. Consequently, the agreement with the reference solution is still very good for $e$, $r$ and $a$, since in those cases the secular evolution of the mean values cancels. On the other hand, the values of $\gamma$ and $a$ given by the regular expansion not only separate very soon from the reference solution, but are also incapable of reproducing the amplitude of the oscillations. The figure for $q_{2}(\theta)$ is particularly interesting, clearly showing the sinusoidal evolution of the mean value of this parameter in the slow scale.

A case of initially circular orbit is considered in Fig.~\ref{fig:radial_e0_eps002}, for a non-dimensional perturbing acceleration of $\varepsilon=0.02$. This example is of special interest for the study of the multiple scales solution, since the expressions used for $g_{1}$ and $g_{2}$ are now exact. It is observed that the amplitudes of the oscillations in both the regular and multiple scales solutions are slightly smaller than the amplitudes for the reference solution; this error is due to the order of the asymptotic approximation, and could be reduced by retaining terms of higher order in the expansion. Regarding the period of the oscillations, the results for initially circular orbit developed in the previous section suggested that it is a combination of the characteristic periods for the fast and slow scales. This behavior is confirmed by the excellent agreement between the oscillation periods of both the reference and the multiple scales solutions, with a small drift driven by the terms of the slow `time' scale neglected in the expansion of $\Omega$. Meanwhile, the regular expansion, which does not take into account the slow `time' scale, shows a slightly shorter period than the order two.

The previous test cases have been chosen to characterize and compare the behavior of both asymptotic solutions from a mathematical point of view. For this purpose, the use of the non-dimensional parameter $\varepsilon$ is most convenient. Nevertheless, it is also interesting to relate these non-dimensional acceleration parameters to some practical scenarios. If we consider a heliocentric reference orbit with $R_0 = 1\,\mathrm{AU}$, the non-dimensional accelerations $\varepsilon=0.005$ and $0.02$ correspond to $A=0.0296\,\mathrm{mm}/\mathrm{s}^2$ and  $0.1186\,\mathrm{mm}/\mathrm{s}^2$, respectively. Taking the total mass of $4100\,\mathrm{kg}$ for the recently launched BepiColombo mission on route to Mercury\footnote{https://www.esa.int/Our\_Activities/Space\_Science/BepiColombo/BepiColombo\_factsheet [accessed 25 June 2019]}, the required thrusts would be of $F=121.6\,\mathrm{mN}$ and $486.3\,\mathrm{mN}$, respectively. The first value is close to the $145\,\mathrm{mN}$ qualified by ESA for the QnetiQ thruster used in the mission [see \cite{clark2013bepicolombo}], whereas the second one exceeds in a factor of two the maximum achievable thrust according to QnetiQ [see \cite{hutchins2015qinetiq}]. On the other hand, if we consider an Earth-bound reference orbit with $R_0=42164\,\mathrm{km}$ (i.e. in the GEO region), the dimensional accelerations would be $1.11\,\mathrm{mm/s}^2$ and $4.5\,\mathrm{mm/s}^2$, respectively. The resulting thrusts for a typical GEO satellite weight in the thousands of kilograms would exceed the current capabilities of low-thrust propulsion systems, but this is not an issue regarding the applicability of the asymptotic expansions; in fact, the smaller the $\varepsilon$ the better the approximation, as shown by the numerical results presented in this section.

\begin{figure*}
\centering \includegraphics[width=0.48\columnwidth]{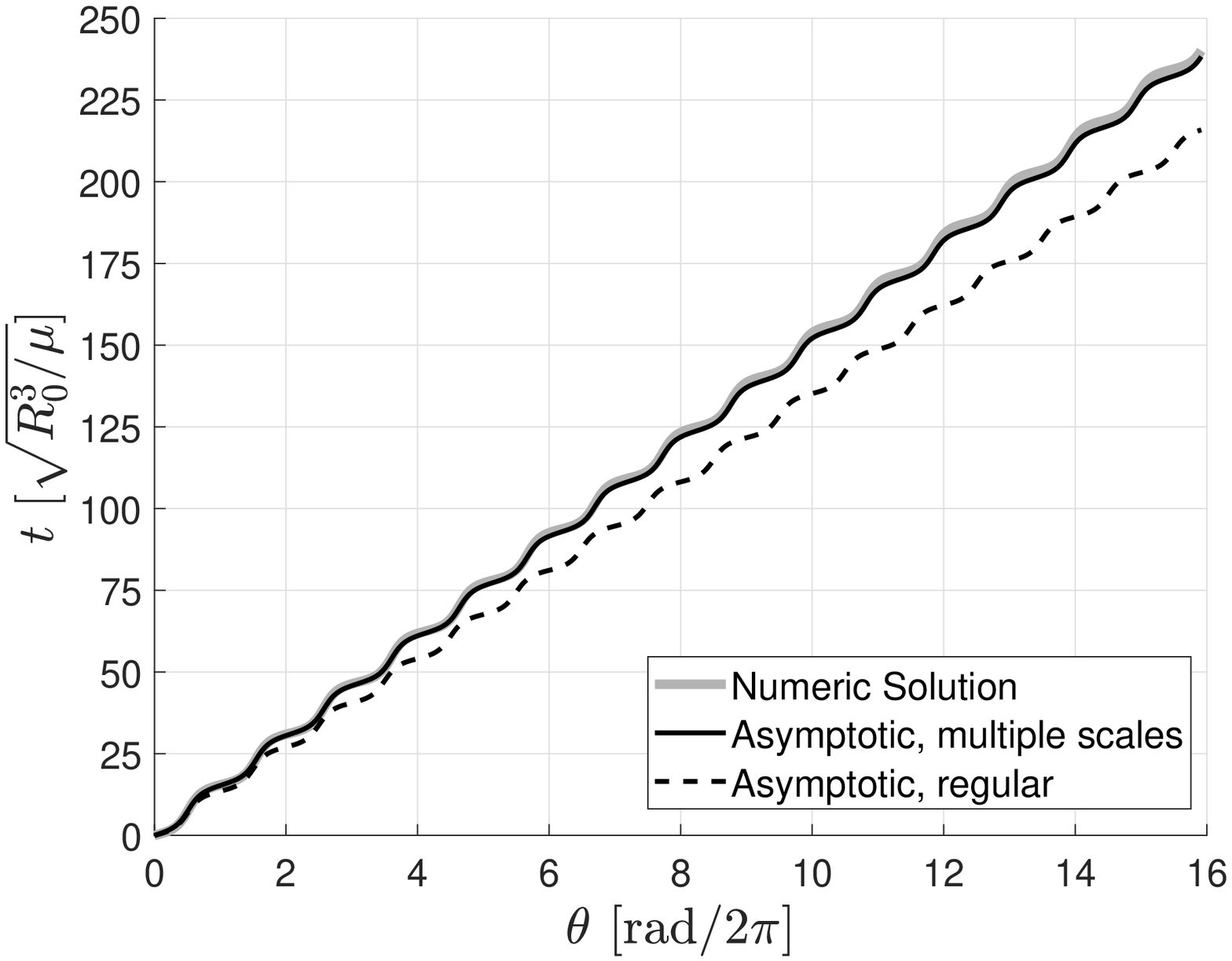} \hfill
\includegraphics[width=0.48\columnwidth]{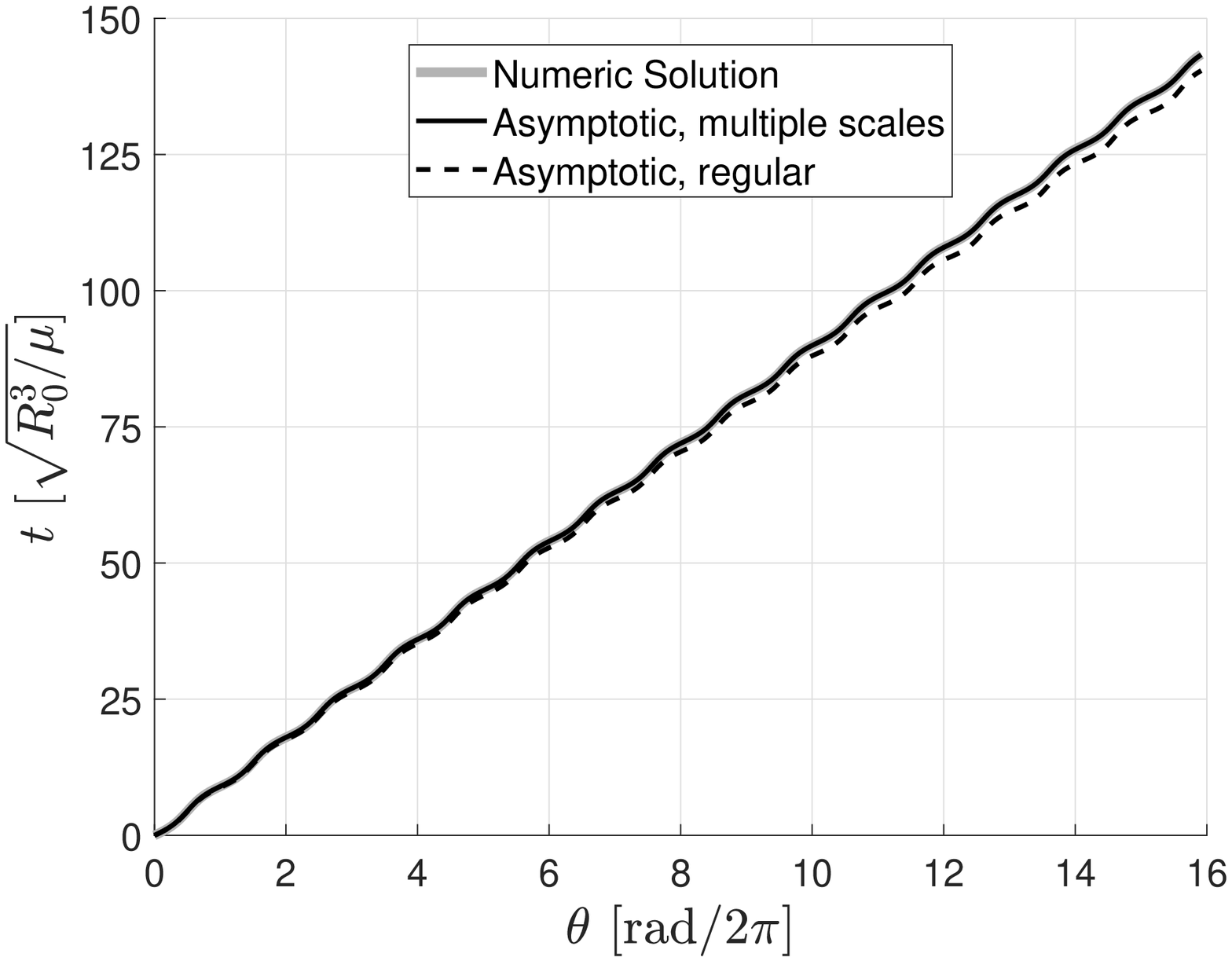}
\caption{Comparison of the non-dimensional times obtained for $e_{0}=0.4$ and $\varepsilon=0.01$ (left), and $e_{0}=0.2$ and $\varepsilon=0.005$ (right).}
\label{fig:radial_time} 
\end{figure*}

The different approximations obtained for the physical time are compared in Fig.~\ref{fig:radial_time}, including the results from a high-precision numerical propagator as reference. It is straightforward to check the good agreement between the numeric (exact) solution and the multiple scales asymptotic solution.

\begin{figure*}
  \centering 
  \includegraphics[width=0.48\textwidth]{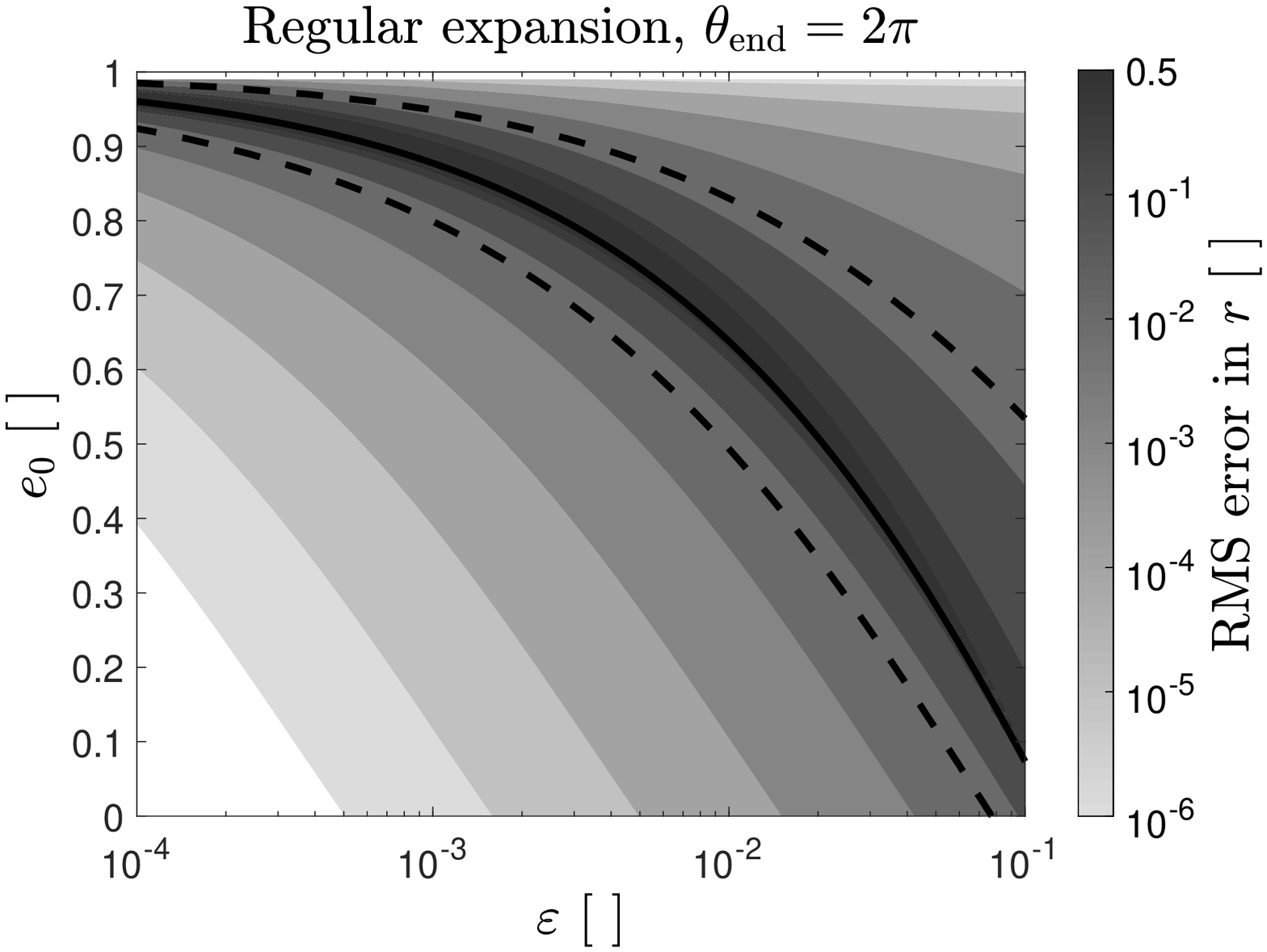}\hfill
  \includegraphics[width=0.48\textwidth]{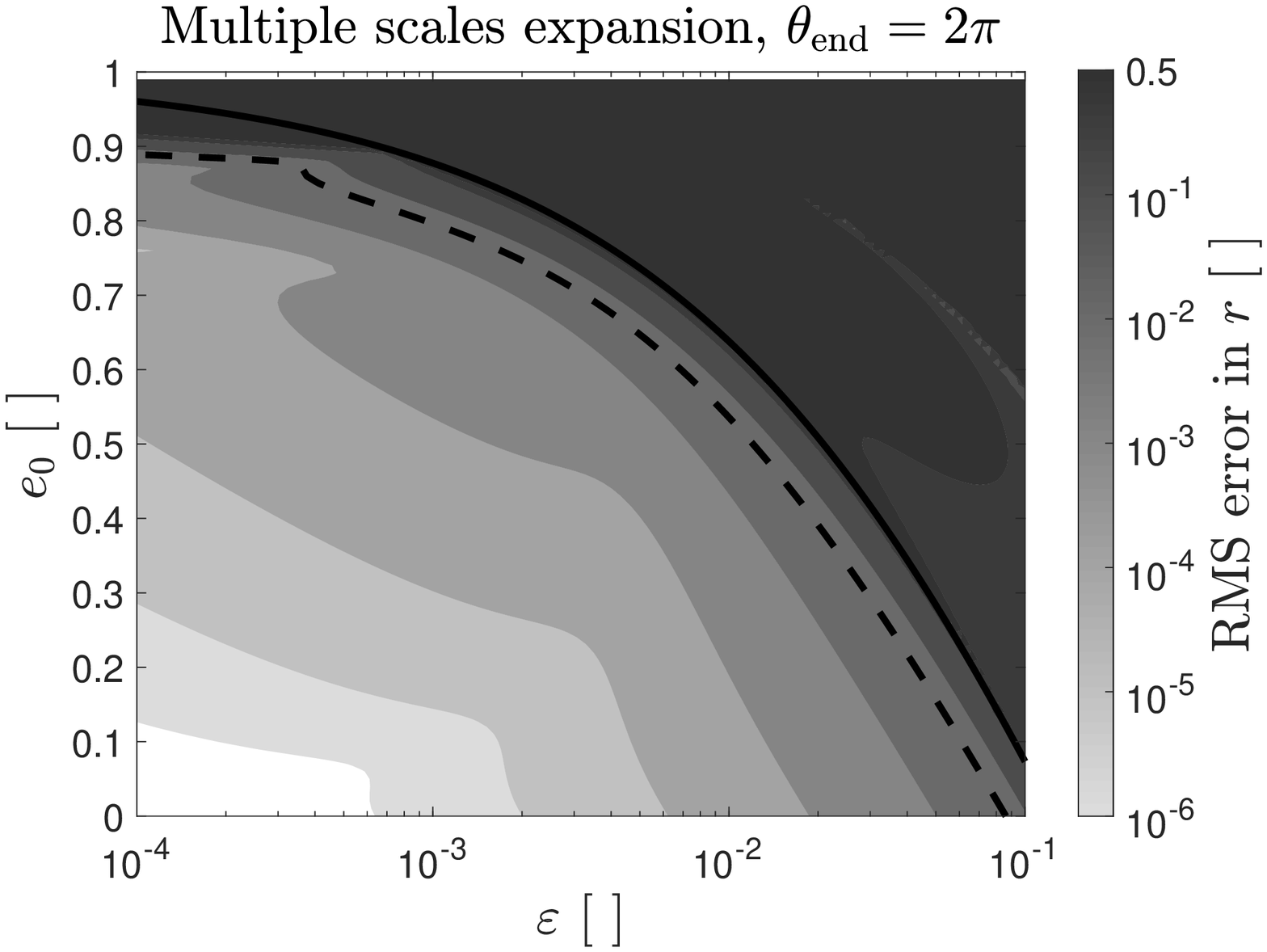} \\[0.8em]
  \includegraphics[width=0.48\textwidth]{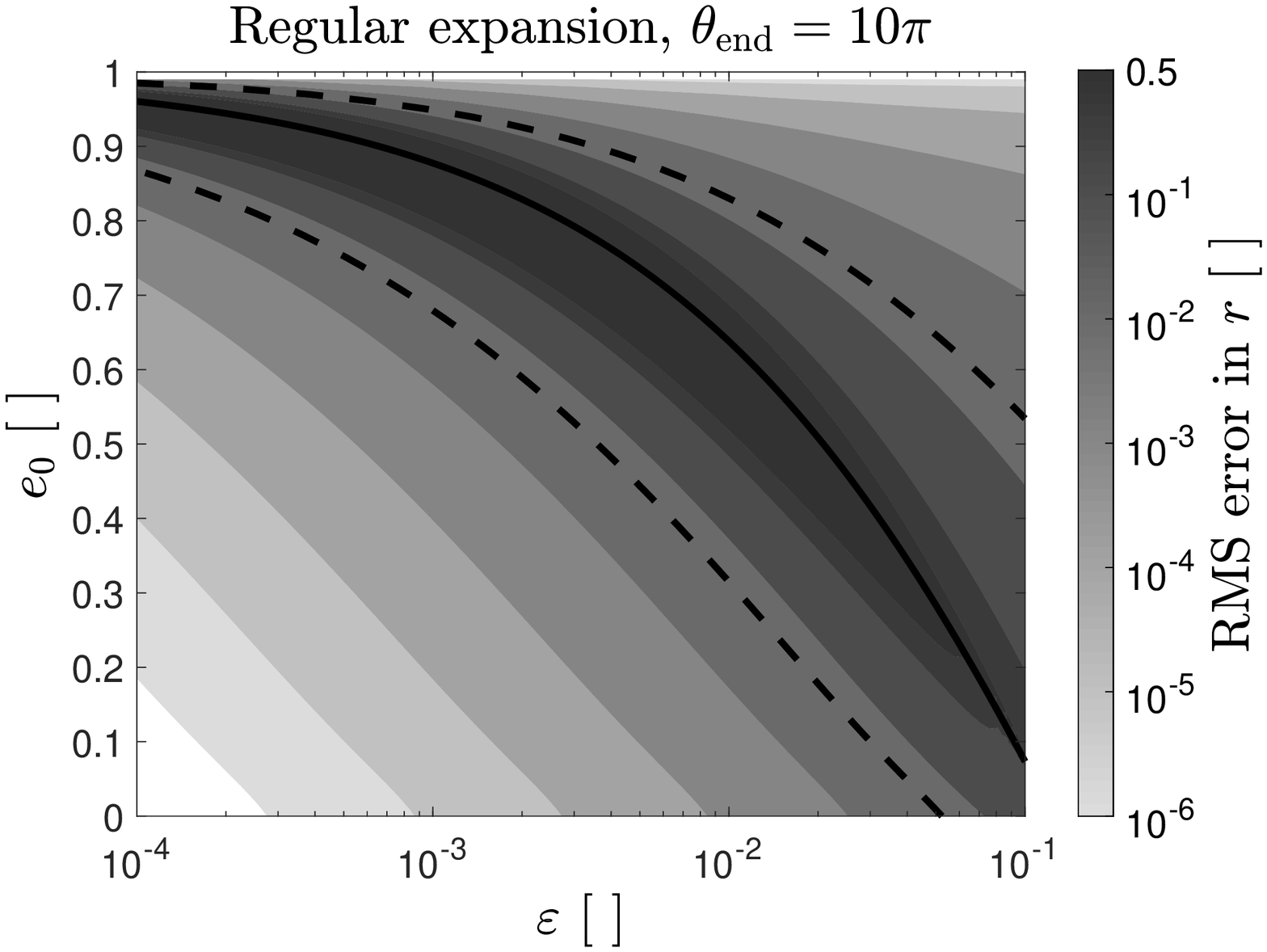}\hfill
  \includegraphics[width=0.48\textwidth]{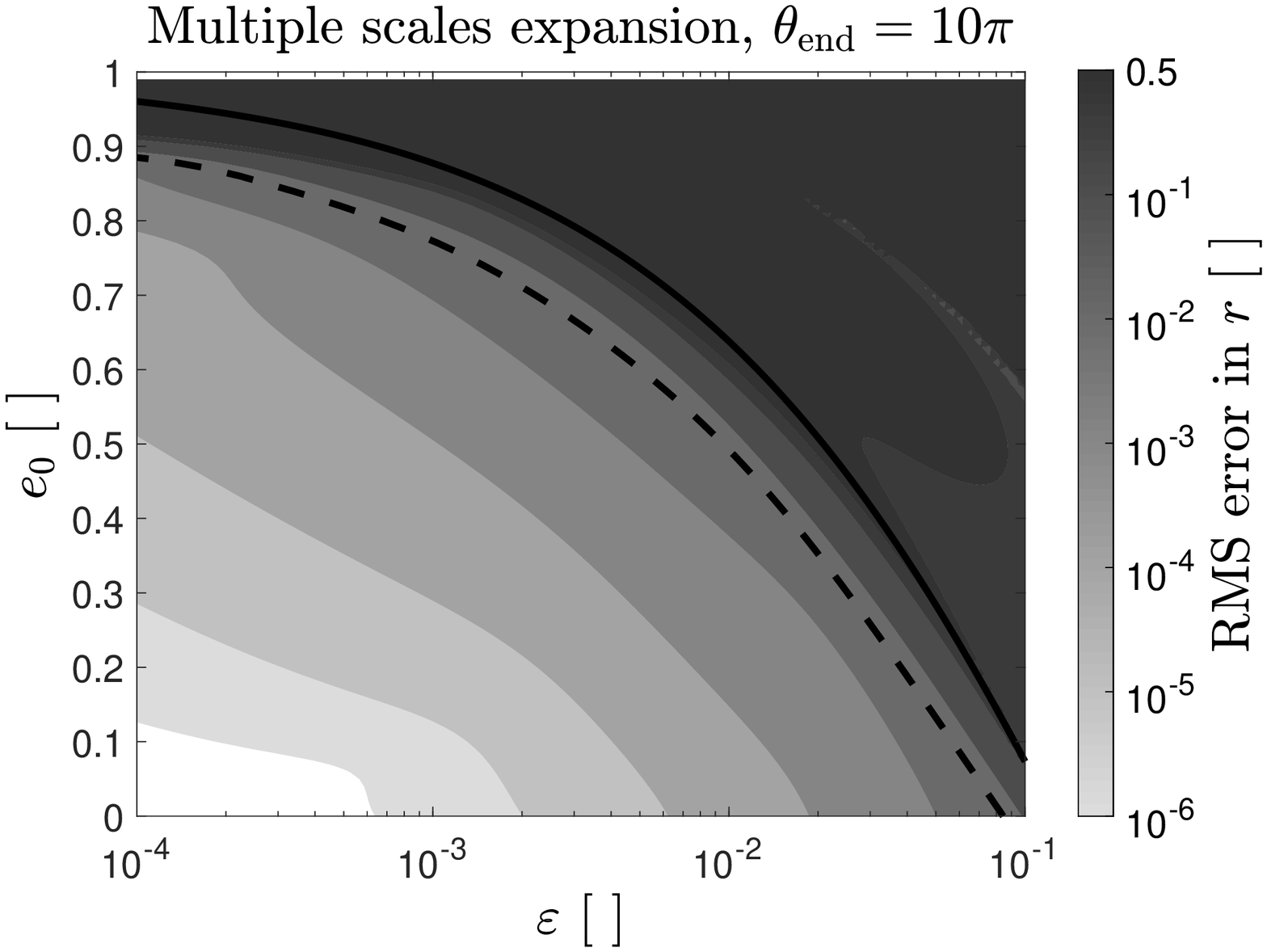} \\[0.8em]
  \includegraphics[width=0.48\textwidth]{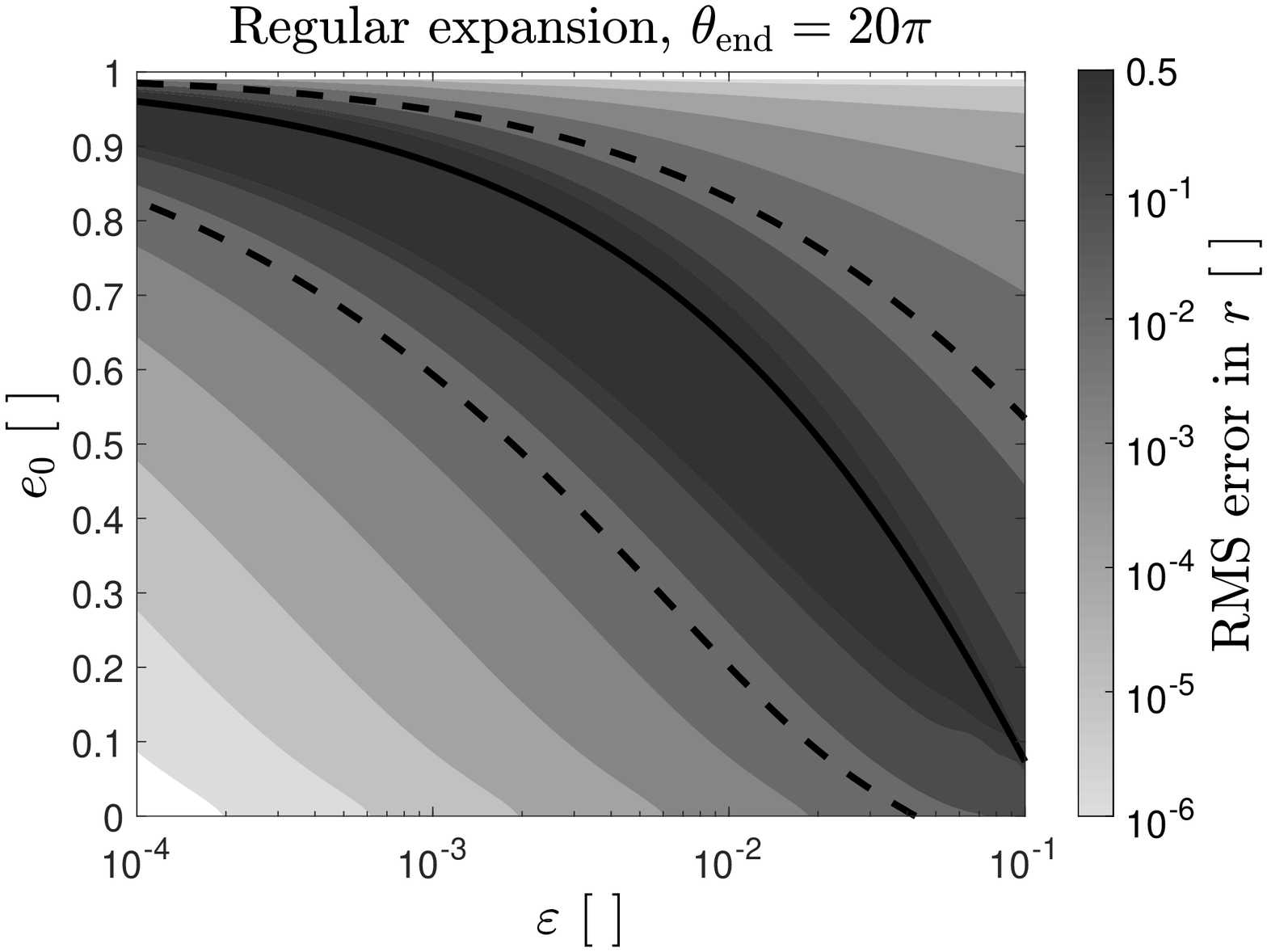}\hfill
  \includegraphics[width=0.48\textwidth]{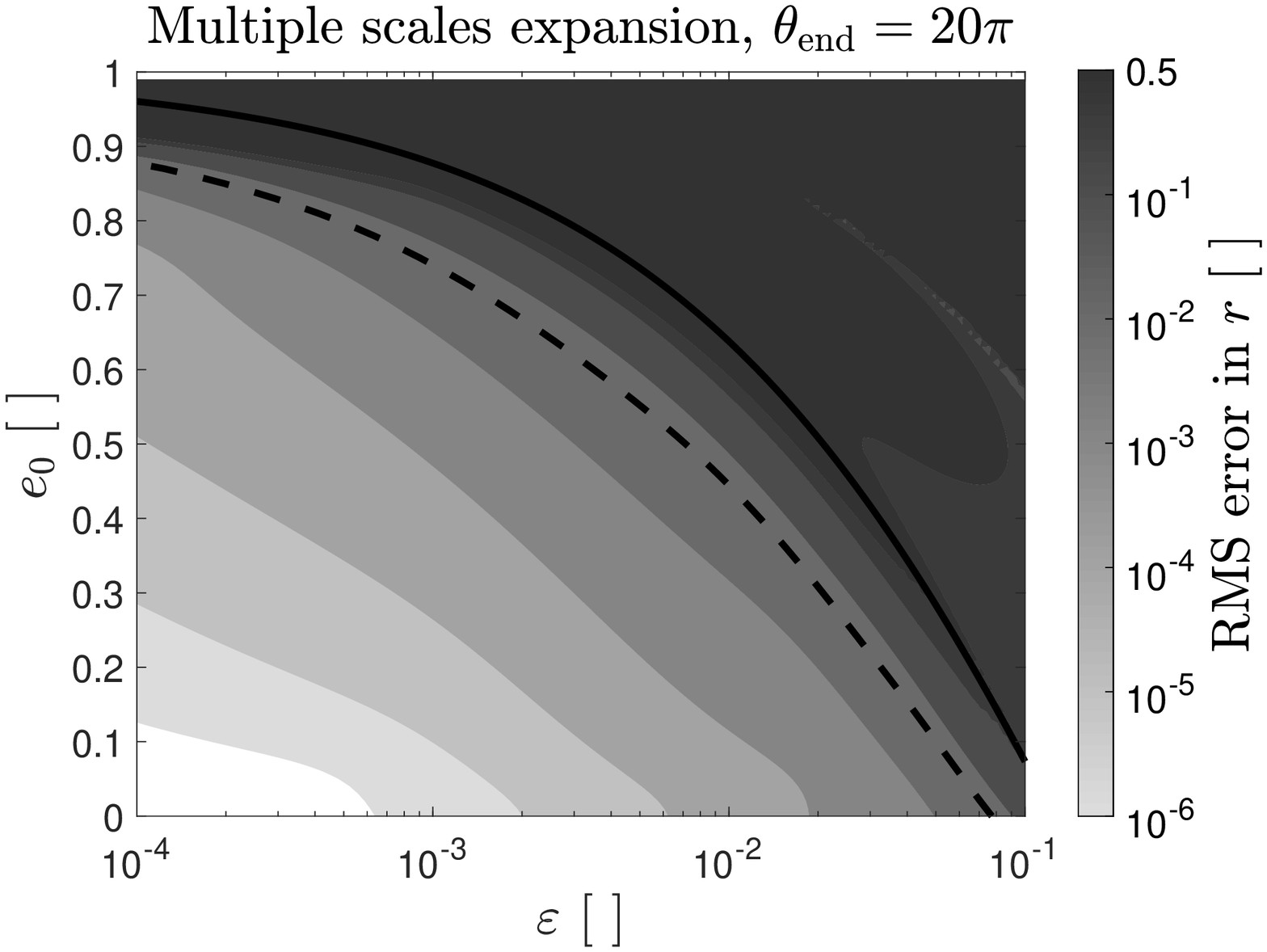} \\[0.8em]
  \includegraphics[width=0.48\textwidth]{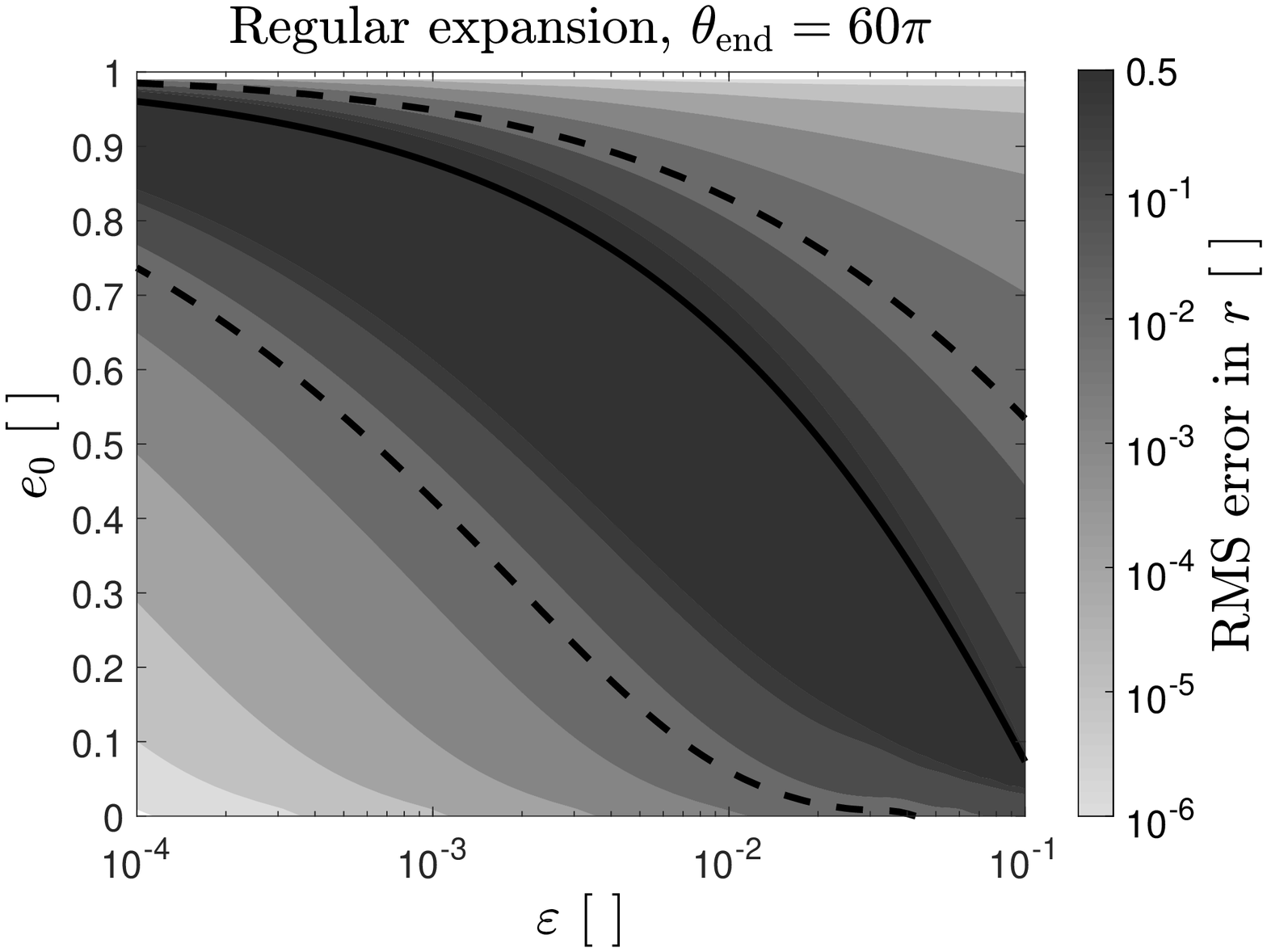}\hfill
  \includegraphics[width=0.48\textwidth]{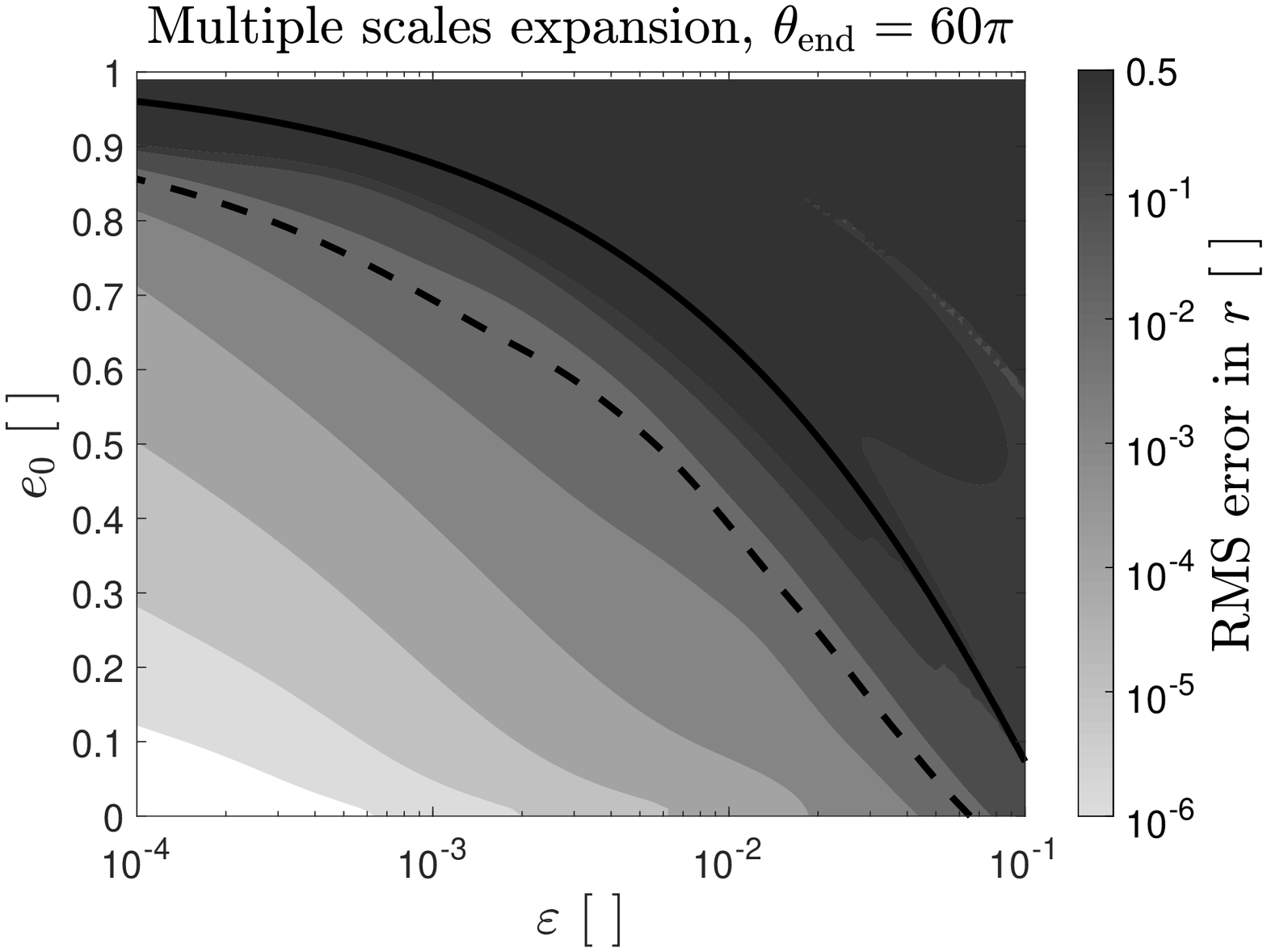}
   \caption{RMS errors in $r$ for the regular and multiple scales asymptotic solutions, as a function of $e_{0}$ and $\varepsilon$. White areas correspond to relative errors lower than $10^{-6}$ and dashed lines to $5\%$ error. The region above the solid black line corresponds to cases reaching escape conditions.}\label{fig:RMSe_r_reg_ms} 
\end{figure*}

The particular cases considered so far have shown that the multiple scales solution behaves much better than the regular expansion, giving a more accurate description of the physics of the problem. Nevertheless, a systematic evaluation of the error of both methods is advisable. To this end, the Root Mean Square (RMS) error in the non-dimensional orbital position has been calculated for a significant range in both the initial eccentricity and the non-dimensional thrust parameter. The results are displayed in Fig.~\ref{fig:RMSe_r_reg_ms}, for several values of the final independent variable $\theta_\mathrm{end}$. In those cases where escape is reached, the error is calculated at the value of $\theta$ for which the energy becomes zero in the numerical integration; a solid line delimits the area corresponding to this group of orbits, associated with high values of $e_{0}$ and $\varepsilon$. Interestingly, the regular expansion behaves better than the multiple scales approximation for cases reaching escape conditions. Because escape takes place during the first orbital revolution the more accurate description of the long period given by the multiple scales solution provides no advantage, while the additional expansions in $e_0$ required for the determination of $g_{1}$, $g_{2}$ and $\Omega_2$ have a negative impact in accuracy. The situation is reversed for bounded orbits, with the multiple scales solution clearly outperforming the regular expansion. Comparing the plots for increasing $\theta_\mathrm{end}$ it is checked that the RMS error for the multiple scales expansion grows very slowly, whereas the accuracy of the regular expansion degrades very fast. This supports the conclusion that the method of multiple scales extends the validity range for the expansion. Looking at the results for $\theta_\mathrm{end}=2\pi$ and small $\varepsilon$, it is possible to identify again the additional error introduced in the multiple scales solution by the expansion in $e_0$ used for $g_{1}$, $g_{2}$ and $\Omega_2$. Note that this additional expansion is not a characteristic feature of the method of multiple scales, but a limitation due to the impossibility of finding an exact solution for those functions.

\begin{figure}
\centering \includegraphics[width=0.6\columnwidth]{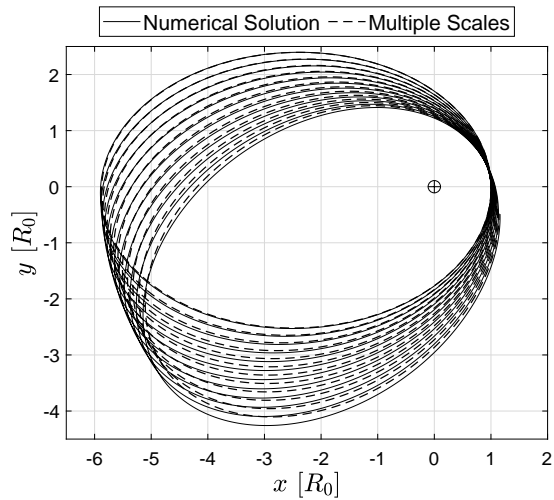}
\caption{Evolution of an orbit with $e_{0}=0.7$ and $\varepsilon=0.001$.}\label{fig:orbits_e07_eps0001} 
\end{figure}

Figure \ref{fig:orbits_e07_eps0001} shows the evolution of an orbit with $e_{0}=0.7$ and $\varepsilon=0.001$, for both the reference and the multiple scales solutions. The asymptotic solution remains close to the exact one during the first orbital revolutions, but slowly separates for larger values of $\theta$. The rotation of the eccentricity vector, given by $\gamma$, can be clearly appreciated.

\begin{figure}
\centering \includegraphics[width=0.6\columnwidth]{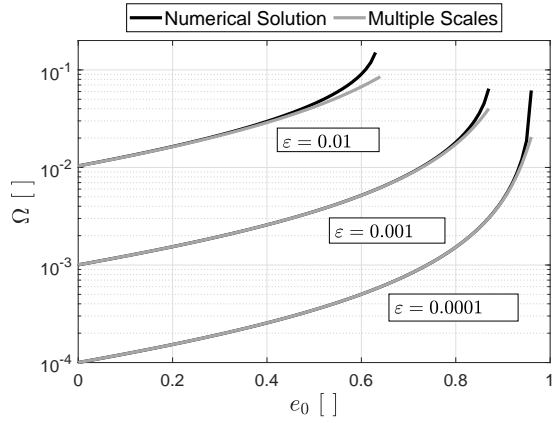}
\caption{Comparison of the characteristic frequency of the long-term evolution of the orbit, obtained using high-precision numerical propagation and the multiple scales asymptotic solution.}\label{fig:Omega_e0_eps} 
\end{figure}

\begin{figure}
\centering \includegraphics[width=0.6\columnwidth]{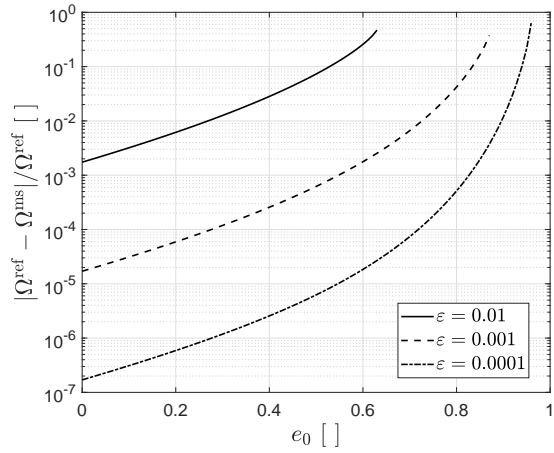}
\caption{Relative error in the characteristic frequency of the long-term evolution of the orbit given by the multiple scales solution, using a high-precision numerical propagation as reference.}\label{fig:Error_Omega_e0_eps} 
\end{figure}

Finally, Figs.~\ref{fig:Omega_e0_eps} and~\ref{fig:Error_Omega_e0_eps} compare the characteristic frequency of the slow scale $\Omega$ given by the multiple scales approximation with a high-precision numerical solution, for different values of $e_0$ and $\varepsilon$. The curves corresponding to each $\varepsilon$ are computed until the $e_0$ leading to escape conditions is reached. As expected, the asymptotic solution is very close to the reference value for small $e_0$ and $\varepsilon$, and progressively separates from it as the initial eccentricity and the acceleration parameter increase. Furthermore, the error decreases quadratically with $\varepsilon$ for moderate values of the initial eccentricity.

\section{Comparison with other solutions}

The constant radial thrust problem, or Tsien problem, has been investigated by several authors along the years, as summarized in the Introduction. In this final section, a brief comparison of the newly proposed multiple scales asymptotic solution with other approaches is provided.

One common characteristic of all the exact solutions, either for the general problem [see \cite{izzo2014explicit}] or for parts of it like time of flight determination, is that they rely on the use of elliptic functions. The Hamiltonian-based solution by \cite{san2012bounded} shows that the three kinds of Jacobi elliptic integrals are intrinsic to the problem. This is consistent with the solution for the most general case in terms of Weierstrass's elliptic and related functions by \cite{izzo2014explicit}, as they can be expressed as a combination of the three Jacobi elliptic functions. One key aspect of \cite{izzo2014explicit} is that they provide a solution for the state as a function of time (explicit solution), a feature also shared by the slightly more recent, Dromo-based work by \cite{urrutxua2015dromo}, whether previous exact solutions only provided the time as a function of the state (implicit). Some interesting applications of exact solutions are determining the conditions for bounded motion or periodic motion, a task that cannot be performed with approximate approaches like the one proposed in this work.

However, \cite{quarta2012new} highlight a number of practical limitations to these solutions when introducing their Fourier-based approximate model. On the one hand, they mention the time-implicit nature of exact solutions (note that the time-explicit exact solutions by \cite{izzo2014explicit} and \cite{urrutxua2015dromo} are posterior to the work by Quarta and Mengali). On the other hand, they stress how the use of elliptic functions hinders their practical application for mission design due to the lack of physical insight and more complex analytical manipulation. The lack of physical insight is also considered by \cite{san2012bounded}, where a qualitative description of the flow in the energy-momentum plane is used to tackle the issue. The asymptotic solution presented in this work successfully addresses these limitations, by providing a explicit result in the fictitious time introduced by the Sundman transformation, and expressing it through familiar circular functions. Moreover, compared to the results by \cite{quarta2012new} it has the advantage of explicitly retaining the two characteristic frequencies of the problem, conveniently separating them into one related to the orbital period and another one scaled by the magnitude of the perturbing acceleration.

It is important to note that the aforementioned practical limitations in the use of elliptic and related functions refer to the lack of physical insight and more difficult analytic manipulation, not the computational cost. Modern algorithms allow for the numerical evaluation of elliptical functions at a complexity not significantly higher than trigonometric functions [for a discussion on this topic see \cite{martinusi2011solutions}].

Finally, it is interesting to take a closer look at Hamiltonian-based methods, as Hamiltonian perturbation techniques are one of the most common approaches in astrodynamics to search for analytical solutions valid for long time scales (as compared to the infrequent use of multiple scales methods). The work by \cite{san2012bounded} proposes a solution for the Tsien problem based on a Hamiltonian formulation in polar coordinates. Because the conjugate momentum of the angular coordinate is an integral of the motion (more precisely, the angular momentum), the flow is separable and a Hamiltonian with one degree of freedom is obtained. Furthermore, given that the Hamiltonian is constant due to the conservation of energy, the problem is integrable (albeit requiring elliptic functions). Through an appropriate choice of the length and time units they manage to remove both $\mu$ and $\varepsilon$ from the equations, thus obtaining a solution valid for all cases. An analysis of the flow in the energy-momentum plane reveals that the separation of regions corresponding to bounded and unbounded motions can be described through a perturbation phenomenon. For the bounded motion case, a full family of canonical transformations leading to a Hamiltonian depending only on the momenta is derived from the Hamilton-Jacobi equation. These transformations can be expressed in terms of the Jacobi elliptic functions of the three kinds, thus giving an analytical solution for the problem. It is important to note that \cite{san2012bounded} don't need to apply perturbation methods with their Hamiltonian formulation, as the problem is fully integrable under the proper choice of a canonical transformation. It would be possible to search for an asymptotic solution avoiding the use of special functions, but this lies beyond the scope of this paper.

\section{Conclusion}

Two asymptotic solutions for the two body problem perturbed by a small, constant acceleration oriented along the radial direction have been obtained, using both regular expansions and the method of multiple scales. The equations of motion have been expressed using the Dromo orbital formulation, which has proven very adequate for this purpose. Some important conclusions are reached from studying the structure and behavior of the regular and multiple scales expansions:
\begin{itemize}
\item The regular expansion fails for $\theta\sim1/\varepsilon$, where $\theta$ is Dromo independent variable, related with the true anomaly, and $\varepsilon$ is the non-dimensional perturbing acceleration. Consequently, it cannot be used to propagate orbits for long periods of time, unless a reinitialization process like the one proposed by \cite{bombardelli2011asymptotic} is included. 
\item The method of multiple scales reveals that the problem has \emph{two fundamental scales}. The first one is responsible of the $2\pi$-periodic oscillations along each orbit; while the second one, with a period depending on $\varepsilon$ and $e_{0}$, drives the long term variations of the mean values and the amplitudes of the oscillations. While the existence of these two scales is a known fact in astrodynamics the use of the multiple scales technique applied to a regularized orbital formulation allows a straightforward computation of the main frequencies governing the dynamics without the employment of special functions.
\item To close the multiple scales solution for the terms of $\mathcal{O}(\varepsilon)$, an additional expansion in the initial eccentricity $e_{0}$ has been introduced. Although its solution is exact for initially circular orbit, it contributes to the degradation of the multiple scales expansion for orbits with higher initial eccentricity.
\item The numerical test cases show that the multiple scales solution behaves noticeably better than the regular expansion in most cases, with a substantially larger expanding interval. One key factor has been the use of a high order method with a coordinate strain function for the slow `time' scale, improving the mathematical description of the long period.
\end{itemize}
Compared to other solutions for the constant radial thrust problem, the multiple scales expansion in Dromo variables has the advantages of providing a physically intuitive separation of the characteristic periods and not requiring the evaluation of special functions. Furthermore, it is interesting to highlight that the mathematical structure of the method of multiple scales identified the existence of just two fundamental periods without requiring additional information.

Finally, the good results obtained from the application of the method of multiple scales confirm its suitability as a mathematical tool for the analysis of problems in orbital mechanics where two (or more) different time scales are clearly identifiable, similarly to other perturbation methods such as averaging. The present work can be potentially used as a starting point for the 
application of the method of multiple scales to other astrodynamics problems in the future.

\appendix

\section{Components of the first order terms of the multiple scales solution}\label{ap:ms_components}

The first order terms $q_{11}$ and $q_{21}$ of the multiple scales solution are given as a combination of the following functions: 
\[
\mathcal{P}_{1}(\tau,T)=-\frac{\left(q_{10}+q_{3i}\right)\left(1+\cos\tau\right)+q_{20}\sin\tau}{q_{3i}\left(q_{3i}^{2}-q_{10}^{2}-q_{20}^{2}\right)s_{0}} \, ,
\]
\[
\mathcal{S}_{1}(T)=-\frac{2q_{20}}{q_{3i}\left(q_{3i}^{2}-q_{10}^{2}-q_{20}^{2}\right)^{3/2}} \, ,
\]
\[
\mathcal{P}_{2}(\tau,T)=\frac{q_{10}q_{20}\left(1+\cos\tau\right)+\left(-q_{3i}^{2}+q_{20}^{2}+q_{3i}q_{10}\right)\sin\tau}{q_{3i}\left(q_{3i}-q_{10}\right)\left(q_{3i}^{2}-q_{10}^{2}-q_{20}^{2}\right)s_{0}} \, ,
\]
\[
\mathcal{S}_{2}(T)=\frac{2q_{10}}{q_{3i}\left(q_{3i}^{2}-q_{10}^{2}-q_{20}^{2}\right)^{3/2}} \, ,
\]
\[
\mathcal{K}(\tau,T)=-\frac{\left(\sqrt{q_{3i}^{2}-q_{10}^{2}-q_{20}^{2}}-q_{3i}+q_{10}\right)\sin\tau-q_{20}(1+\cos\tau)}{s_{0}-q_{10}-q_{3i}\cos\tau+(1+\cos\tau)\sqrt{q_{3i}^{2}-q_{10}^{2}-q_{20}^{2}}} \, ,
\]
where $q_{10}$ and $q_{20}$ are functions only of the slow `time' scale $T$.

Introducing the first integral $q_{10}^{2}+q_{20}^{2}=q_{1i}^{2}$, these expressions take simpler forms:
\begin{equation}\label{eq:radial_ms_P11}
\mathcal{P}_{1}(\tau,T)=-\frac{\left(q_{10}+q_{3i}\right)\left(1+\cos\tau\right)+q_{20}\sin\tau}{q_{3i}\left(q_{3i}^{2}-q_{1i}^{2}\right)s_{0}} \, ,
\end{equation}
\begin{equation}\label{eq:radial_ms_S11}
\mathcal{S}_{1}(T)=-\frac{2q_{20}}{q_{3i}\left(q_{3i}^{2}-q_{1i}^{2}\right)^{3/2}}=-2\Omega_{1}q_{20} \, ,
\end{equation}
\begin{equation}\label{eq:radial_ms_P21}
\mathcal{P}_{2}(\tau,T)=\frac{q_{10}q_{20}\left(1+\cos\tau\right)+\left(-q_{3i}^{2}+q_{20}^{2}+q_{3i}q_{10}\right)\sin\tau}{q_{3i}\left(q_{3i}-q_{10}\right)\left(q_{3i}^{2}-q_{1i}^{2}\right)s_{0}} \, ,
\end{equation}
\begin{equation}\label{eq:radial_ms_S21}
\mathcal{S}_{2}(\tau,T)=\frac{2q_{10}}{q_{3i}\left(q_{3i}^{2}-q_{1i}^{2}\right)^{3/2}}=2\Omega_{1}q_{10} \, ,
\end{equation}
\begin{equation}\label{eq:radial_ms_K}
\mathcal{K}(\tau,T)=-\frac{\left(\sqrt{q_{3i}^{2}-q_{1i}^{2}}-q_{3i}+q_{10}\right)\sin\tau-q_{20}(1+\cos\tau)}{s_{0}-q_{10}-q_{3i}\cos\tau+(1+\cos\tau)\sqrt{q_{3i}^{2}-q_{1i}^{2}}} \, .
\end{equation}

\section{Expressions for $\Omega_{2}$, $g_{1}(T)$ and $g_{2}(T)$}\label{ap:g_functions}

As stated in Section \ref{sec:MultipleScales}, the direct application of the secularity condition which determines the values of $\Omega_{2}$, $g_{1}(T)$ and $g_{2}(T)$ is not feasible due to the complexity of the equations involved. To address this issue, an approximate solution is sought for in the form of an additional series expansion in the initial value of $q_{1}$, related to the initial eccentricity $e_{0}$. Then, for $q_{1i}\ll1$ (i.e. $e_0 \ll 1$) it is possible to write:
\[
\Omega_{2}=\Omega_{2}^{(0)}+q_{1i}\Omega_{2}^{(1)}+q_{1i}^{2}\Omega_{2}^{(2)}+\ldots \, ,
\]
\begin{align*}
g_{1}(T;q_{1i})=g_{1}^{(0)}(T)+q_{1i}g_{1}^{(1)}(T)+q_{1i}^{2}g_{1}^{(2)}(T)+\ldots \, , \\
g_{2}(T;q_{1i})=g_{2}^{(0)}(T)+q_{1i}g_{2}^{(1)}(T)+q_{1i}^{2}g_{2}^{(2)}(T)+\ldots \, .
\end{align*}
The initial conditions for $g_{1}$ and $g_{2}$ are obtained from Eq.~\eqref{eq:Dromo_radial_ms_q1q2} knowing that $q_{11}(0)=q_{21}(0)=0$:
\[
g_{1}(0)=\frac{2}{q_{3i}\left(q_{3i}^{2}-q_{1i}^{2}\right)}=g_{1i}\,,\quad g_{2}(0)=0\,.
\]
To achieve a better reproduction of the physics of the problem, $\Omega_{1}$ is retained as fixed parameter instead of $q_{3i}$. The latter is then expanded as a power series of $q_{1i}$ using the definition of $\Omega_{1}$, Eq.~\eqref{eq:Omega1_q10_q20}: 
\[
q_{3i}=\Omega_{1}^{-1/4}+q_{1i}^{2}\frac{3}{8}\Omega_{1}^{1/4}-q_{1i}^{4}\frac{3}{128}\Omega_{1}^{3/4}-q_{1i}^{6}\frac{7}{1024}\Omega_{1}^{5/4}+q_{1i}^{8}\frac{195}{32768}\Omega_{1}^{7/4}+\mathcal{O}\left(q_{1i}^{10}\right)\,.
\]
Substituting back into $g_{1i}$ and expanding: 
\[
g_{1i}=2\Omega_{1}^{3/4}-q_{1i}^{2}\frac{1}{4}\Omega_{1}^{5/4}+q_{1i}^{4}\frac{5}{64}\Omega_{1}^{7/4}-q_{1i}^{6}\frac{11}{512}\Omega_{1}^{9/4}+q_{1i}^{8}\frac{51}{16384}\Omega_{1}^{11/4}+\mathcal{O}\left(q_{1i}^{10}\right) \, .
\]
Introducing the previous expressions into Eq.~\eqref{eq:Dromo_radial_ms_o2}, expanding in power series of $q_{1i}$ and imposing the secularity condition to each of the ODE systems obtained from canceling the terms of equal powers of $q_{1i}$, a sequence of straightforward problems for determining $\Omega_{2}^{(k)}$, $g_{1}^{(k)}$ and $g_{2}^{(k)}$ is reached. Note that, due to the structure of these problems, the asymptotic solution for $\Omega_{2}$ will have one order less than those for $g_{1}$ and $g_{2}$. The coefficients obtained for $\Omega_{2}$ up to order $7$ are: 
\[
\left\lbrace \Omega_{2}^{(k)}\right\rbrace =\left\lbrace \frac{7}{2}\Omega_{1}\:,\;3\Omega_{1}^{5/4}\:,\;\frac{9}{4}\Omega_{1}^{3/2}\:,\;\frac{3}{8}\Omega_{1}^{7/4}\:,\;0\:,\;-\frac{9}{128}\Omega_{1}^{9/4}\:,\;-\frac{9}{128}\Omega_{1}^{5/2}\:,\;\frac{9}{1024}\Omega_{1}^{11/4}\right\rbrace \quad k=0,\ldots,7\,.
\]
On the other hand, the first two terms of $g_{1}$ and $g_{2}$ take the form: 
\[
g_{1}^{(0)}(T)=\Omega_{1}^{3/4}\left(1+\cos T\right) \, , \quad g_{2}^{(0)}(T)=\Omega_{1}^{3/4}\sin T \, ,
\]
\[
g_{1}^{(1)}(T)=0\,,\quad g_{2}^{(1)}(T)=-\Omega_{1}\sin T \, .
\]
For the rest of terms, at least up to order eight, a simple functional expression can be found:
\[
g_{1}^{(k)}(T)=\Omega_{1}^{\frac{3+k}{4}}\left[C_{0}^{(k)}+\sum_{l=1}^{k}C_{l}^{(k)}\cos lT\right] \, ,
\]
\[
g_{2}^{(k)}(T)=-\deriv{g_{1}^{(k)}}{T}=\Omega_{1}^{\frac{3+k}{4}}\sum_{l=1}^{k}lC_{l}^{(k)}\sin lT \, ,
\]
with coefficients:
\[
\mathbf{C}^{(2)}=-\frac{1}{8}\left[\begin{array}{ccc}
-3 & 1 & 4\end{array}\right] \, ,
\]
\[
\mathbf{C}^{(3)}=\frac{1}{8}\left[\begin{array}{cccc}
0 & 1 & 0 & -1\end{array}\right] \, ,
\]
\[
\mathbf{C}^{(4)}=\frac{1}{384}\left[\begin{array}{ccccc}
-9 & 15 & 40 & 0 & -16\end{array}\right] \, ,
\]
\[
\mathbf{C}^{(5)}=\frac{1}{64}\left[\begin{array}{cccccc}
0 & -2 & 0 & 3 & 0 & -1\end{array}\right] \, ,
\]
\[
\mathbf{C}^{(6)}=\frac{1}{5120}\left[\begin{array}{ccccccc}
-35 & -55 & -100 & 0 & 112 & 0 & -32\end{array}\right] \, ,
\]
\[
\mathbf{C}^{(7)}=\frac{1}{768}\left[\begin{array}{cccccccc}
0 & 3 & 0 & -9 & 0 & 8 & 0 & -2\end{array}\right] \, ,
\]
\[
\mathbf{C}^{(8)}=\frac{1}{229376}\left[\begin{array}{ccccccccc}
1365 & 357 & -336 & 0 & -1568 & 0 & 1152 & 0 & -256\end{array}\right] \, .
\]

This solution is obtained expanding the initial condition for $g_{1}$ as a power series in $q_{1i}$. Alternatively, $g_{1i}$ could be retained as a fixed parameter, leading to a solution which keeps the same structure and most of the coefficients. Particularly, for $\Omega_{2}$ only the odd terms are modified (in the following, the left-hand side represents the new coefficient, and the right-hand side its expression in terms of the old coefficient): 
\begin{gather*}
\Omega_{2}^{(1)}\rightarrow3\Omega_{1}^{1/4}g_{1i}-\Omega_{2}^{(1)}\,,\quad\Omega_{2}^{(3)}\rightarrow\frac{3}{4}\Omega_{1}^{3/4}g_{1i}-\Omega_{2}^{(3)} \, ,\\
\Omega_{2}^{(5)}\rightarrow\frac{3}{32}\Omega_{1}^{3/4}g_{1i}-\Omega_{2}^{(5)}\,,\quad\Omega_{2}^{(7)}\rightarrow-\Omega_{2}^{(7)} \, .
\end{gather*}
Similarly, for $g_{1}$ and $g_{2}$ the only differences are in the zeroth order terms of the expansion: 
\begin{equation}
g_{1}^{(0)}=\left(g_{1i}-\Omega_{1}^{3/4}\right)\cos T+\Omega_{1}^{3/4}\,,\quad g_{2}^{(0)}=\left(g_{1i}-\Omega_{1}^{3/4}\right)\sin T \, ,
\end{equation}
and in the coefficients for the rest of even terms: 
\begin{equation}
C_{1}^{(2)}\rightarrow-C_{1}^{(2)}\,,\quad C_{1}^{(4)}\rightarrow-C_{1}^{(4)}\,,\quad C_{1}^{(6)}\rightarrow-C_{1}^{(6)}\,,\quad C_{1}^{(8)}\rightarrow-C_{1}^{(8)} \, .
\end{equation}

\section{Function $g_t(T)$}\label{ap:gtime_functions}

The additive function in $\zeta_{00}\left(\tau,T\right)$ arising from the secularity condition is approximated through its power series expansion in $q_{1i}$, valid for small initial eccentricity:
\[
g_t(T)=g_t^{(0)}(T)+q_{1i}g_t^{(1)}(T)+q_{1i}^{2}g_t^{(2)}(T)+\ldots \, .
\]
All the resulting terms are monomials of $T$, that is, $g_t^{(k)} = D^{(k)} T$, so the expansion can be expressed in a more convenient form as: 
\[
g_t(T)=\left(D^{(0)}+q_{1i}D^{(1)}+q_{1i}^{2}D^{(2)}+\ldots\right)T=D_t T \, ,
\]
with coefficients up to order seven :
\[
\left\lbrace D^{(k)}\right\rbrace =\left\lbrace 2\Omega_{1}^{3/4}\:,\;3\Omega_{1}\:,\;\frac{15}{4}\Omega_{1}^{5/4}\:,\;\frac{3}{2}\Omega_{1}^{3/2}\:,\;\frac{45}{64}\Omega_{1}^{7/4}\:,\;0\:,\;-\frac{67}{512}\Omega_{1}^{9/4}\:,\;-\frac{3}{64}\Omega_{1}^{5/2}\right\rbrace \quad k=0,\dots,7 \, .
\]

\section{Regular expansion of the physical time}\label{ap:time_regular}

An expression relating the non-dimensional physical time $\hat{t}$ with the fictitious time $\theta$ for the regular expansion solution can be obtained directly from Eq.~\eqref{eq:Dromo_ODEs_time}. To this end, $\hat{t}$ is expressed as a power series of $\varepsilon$:
\begin{equation}
\hat{t}(\theta) = \hat{t}_0(\theta) + \varepsilon \hat{t}_1(\theta) + \varepsilon^2 \hat{t}_2(\theta) + \mathcal{O}\left(\varepsilon^3\right) \, .
\end{equation}
Introducing this expression and the known asymptotic solutions for $\hat{q}_1$ and $\hat{q}_2$ into Eq.~\eqref{eq:Dromo_ODEs_time}, expanding for $\varepsilon \ll 1$ and retaining only the leading order terms, the following differential equation for $\hat{t}_0(\theta)$ is reached:
\begin{equation*}
\deriv{\hat{t}_0}{\theta} = \frac{1}{q_{3i} \left( \hat{s}_0 \right)^2} \, ,
\end{equation*}
which upon integration yields:
\begin{equation}
\hat{t}_0 = \frac{-q_{1i}\sin\theta}{q_{3i}\left( q_{3i}^2-q_{1i}^2\right) \hat{s}_0 } + \frac{1}{\left( q_{3i}^2 - q_{1i}^2 \right)^{3/2}} \left( \theta + \arctan \hat{\mathcal{K}} \right) \, .
\end{equation}
This expression includes both secular and oscillatory terms in $\theta$. Unlike what happened with $\hat{q}_2$, the presence of a secular term was expected and necessary, since time must be monotonically increasing with $\theta$. Also note that the secular term coincides with the solution for the time element in Eq.~\eqref{eq:reg_time_el}.

\begin{acknowledgements}

This work has been supported by the Spanish Ministry of Education, Culture and Sport through its FPU Program (reference number FPU13/05910). It has also received funding from the project ``Dynamical Analysis of Complex Interplanetary Missions'' (ref. ESP2017-87271-P) supported by the Agencia Estatal de Investigación (AEI) of the Spanish Ministry of Economy, Industry and Competitiveness (MINECO) and by the European Fund of Regional Development (FEDER).
The authors also thank the two anonymous reviewers, the associated editor, and Dr. Ioannis Gkolias for their useful comments.

\end{acknowledgements}

\section*{Compliance with ethical standards}
\textbf{Conflict of interest:} The authors declare that they have no conflict of interest.

\bibliographystyle{spbasic}

\end{document}